\documentclass[fleqn,usenatbib]{mnras}

\usepackage[T1]{fontenc}
\usepackage{ae,aecompl}

\usepackage{graphicx}	
\usepackage{amsmath}	
\usepackage{amssymb}	

\title[Stellar populations at $z\sim1.3$]{Age, metallicity and star formation 
history of spheroidal galaxies in cluster at $z\sim1.2$}

\author[P. Saracco et al.]{
P. Saracco $^{1}$\thanks{E-mail: paolo.saracco@inaf.it},
F. La Barbera$^{2}$, A. Gargiulo$^{3}$, F. Mannucci$^{4}$,
D. Marchesini$^{5}$,
\newauthor{ 
M. Nonino$^{6}$, P. Ciliegi$^7$}\\
$^{1}$INAF - Osservatorio Astronomico di Brera, via Brera 28, 20121 Milano, Italy\\
$^{2}$INAF - Osservatorio Astronomico di Capodimonte, sal. Moiariello 16, 80131 Napoli, Italy\\
$^{3}$INAF - Istituto di Astrofisica e Fisica Cosmica, IASF, via E. Bassini 15, 20133 Milano, Italy\\
$^{4}$INAF - Osservatorio Astrofisico di Arcetri, Largo Enrico Fermi 5, 50125 Firenze, Italy\\
$^{5}$Tufts University, Physics and Astronomy Department, 574 Boston Ave, Medford, 02155 MA, USA\\
$^{6}$INAF - Osservatorio Astronomico di Trieste, Via G.B. Tiepolo 11, 34143 Trieste, Italy\\
$^{7}$INAF - Osservatorio Astronomico di Bologna, Via Piero Gobetti, 93/3, 40129 Bologna, Italy
}

\date{Accepted 2018 December 17. Received 2018 November 10; in original form
2018 September 11}

\pubyear{2018}

\begin{document}
\label{firstpage}
\pagerange{\pageref{firstpage}--\pageref{lastpage}}
\maketitle

\begin{abstract}
We present the analysis, based on spectra collected at the
{\it Large Binocular Telescope}, of the stellar populations in seven spheroidal 
galaxies in the cluster XLSSJ0223 at $z$$\sim$1.22.
The aim is to constrain the epoch of their formation
and their star formation history.
Using absorption line strenghts and full spectral fitting, 
we derive for the stellar populations of the seven spheroids
a median age <Age>=2.4$\pm$0.6 Gyr, corresponding to a median formation redshift 
$<$$z_f$$>$$\sim2.6_{-0.5}^{+0.7}$ (lookback time = 11$_{-1.0}^{+0.6}$ Gyr).
We find a significant scatter in age, showing that  
massive spheroids, at least in our targeted cluster, are not coeval.
The median metallicity is [Z/H]=0.09$\pm$0.16,
as for early-types in clusters at 0$<$$z$<0.9.
This lack of evolution of [Z/H] over the range 0$<$$z$$<$1.3,
corresponding to the last 9 billions years, 
suggests that no significant additional star formation and 
chemical enrichment are required for cluster spheroids to reach the present-day 
population. 
We do not detect significant correlation 
between age and velocity dispersion $\sigma_e$, or dynamical mass  M$_{dyn}$, or 
effective stellar mass density $\Sigma_e$.
On the contrary, the metallicity [Z/H] of the seven spheroids is correlated to 
their dynamical mass M$_{dyn}$,
according to a relation similar to the one for local spheroids.
[Z/H] is also anticorrelated to stellar mass 
density $\Sigma_e$ because of the anticorrelation between M$_{dyn}$ and  $\Sigma_e$. 
Therefore, the basic trends observed in the local universe were already 
established at $z\sim1.3$, i.e. more massive spheroids are more 
metal rich, have lower stellar mass density and 
tend to be older than lower-mass galaxies.
\end{abstract}




\section{Introduction}
The mechanisms through which  massive (stellar mass M$_*$$\approx$$10^{11}$ M$_\odot$) 
spheroidal galaxies assemble and shape their stellar mass and then quench 
their star formation are still unclear and represent central
topics in galaxy evolution.
Linking the population of galaxies at different redshifts to constrain 
their evolution, it is not always a reliable technique, since it is affected 
by the well known progenitor bias problem 
\citep[e.g.][]{vandokkum01a,carollo13}.
The study of spheroids, compared to other types of galaxies, is potentially less 
affected by this bias, as once an high-density bulge is formed, it is unlikely 
that it is disrupted, or assembles efficiently a surrounding disc \citep[e.g.][]{brooks16}.
Therefore, even if not all the local spheroidal galaxies may have a spheroid 
as progenitor, high-redshift spheroidal galaxies are most probably the progenitors
of some of the local ones.

Simulations suggest that an early intense burst
of star formation followed by quenching, is required to reproduce 
the detailed structural properties 
of ellipticals and to match the observed tight scaling relations
\citep{ciotti07,Naab07,oser12,porter14a,brooks16}.
Observations show that only a minor fraction of local
spheroids has accreted newly stellar mass through secondary events of 
star formation \cite[e.g.][]{thomas10,gargiulo16}.
Therefore, the stellar populations of high-redshift spheroidal 
galaxies keep the information of the early-phases of their formation. 
If their evolution is mostly characterized by passive ageing, then
their stellar population properties hold on nearly unchanged till now.
Consequently, stellar chemical composition and age are powerful tools to link 
spheroids across time and, most importantly, to constrain their formation 
and their past evolution. 

Correlations between stellar population parameters and structural parameters of 
early-type galaxies have been extensively studied in the past.
The line strengths of early-type galaxies in the optical spectral range are found to be
correlated or anticorrelated to the velocity dispersion, depending whether
the lines are more sensitive to metallicity or age effects 
\cite[e.g.][]{bender93,fisher95,colless99,jorgensen99,trager00,
bernardi03, bernardi06, harrison11, mcdermid15, jorgensen17}.
These relationships are observed both for field and cluster early-type galaxies 
in the local universe and it seems that environment does not affect them significantly
 \citep[e.g.][]{bernardi06,mcdermid15}.
For cluster early-types these relations are observed and well defined up to $z$$\sim$0.9
\citep{jorgensen17}. 

Given the dependence of line-strengths on metallicity and age,
it is expected a correlation between age, metallicity
and velocity dispersion.
Indeed, the metallicity [Z/H] of local spheroidal galaxies, tipically solar or super-solar 
\cite[e.g.][]{buzzoni92,jorgensen99,thomas05},  
is found to correlate with central velocity dispersion 
\citep{greggio97,jorgensen99,trager00,thomas05,thomas10,harrison11,mcdermid15}.
Contrary to metallicity, the correlation between age and velocity 
dispersion is much less evident and it is more uncertain.
Some authors find weak but significant correlation, with older galaxies 
characterized by higher velocity dispersions 
\cite[e.g.][]{thomas05,gallazzi05,gallazzi14,harrison11,mcdermid15}.
Other authors, find shallower correlations and slopes consistent
with zero \cite[e.g.][]{trager00, jorgensen17}, especially when age 
is estimated over an aperture large enough to enclose most of the galaxy 
light \citep{labarbera10a}.
Overall, considering velocity dispersion a proxy of galaxy mass,
the above relations suggest that more massive galaxies are more 
metal rich and, possibly, older than their lower mass counterparts.

Actually, the mass-age and the mass-metallicity relations are observed 
\citep[e.g.][]{thomas10,gallazzi14,mcdermid15} and both appear well defined 
when dynamical mass is considered, with the slope of the mass-age 
relation being rather flat.
Thus, most massive early-type galaxies in the local universe are also the 
most metal rich, which appears counter-intuitive if they are also the oldest 
\citep{greggio11}.
However, this could be explained by a deep gravitational potential 
(i.e. a large dark matter halo) that, on one hand, modulates an intense star formation 
and, on the other hand, efficiently retains the metals rapidly produced by 
high-mass stars.
Another possibility is that, metal rich stars are added later 
to the stellar population through later episodes of star formation. 

Whether and which of the above scenarios is the right one 
can be assessed by studing the stellar population properties of high mass 
spheroids in the past, by establishing whether the relations between stellar 
population properties and physical and dynamical properties were already 
in place early on in the evolution of spheroids or not, and whether
the local age-velocity dispersion and age-mass relations hold on
at high-redshift once evolved back in time.

To investigate these issues, we have performed new spectroscopic observations 
for 14 spheroids candidate
members of the cluster XLSSJ0223-0436 at $z\sim1.22$.
The whole data set will be presented in a forthcoming paper.
Here, we present the analysis of the seven spheroids whose spectra 
have S/N high enough to allow for the study of their stellar population
properties.
The paper is organized as follows.
Section 2 briefly describes the sample, the observations and the data reduction.
Section 3 describes the measurement of redshift and absorption line indices.
In Section 4, we describe the constraints obtained from absorption line strengths on
age and metallicity of the seven spheroids.
In Section 5 we make use of full spectral fitting to derive age and metallicity
of the galaxy stellar populations and to constrain their star formation histories (SFHs).
In Section 6 we study the relationships of age and metallicity with
physical and structural parameters, velocity dispersion, dynamical mass
and stellar mass density.
Section 7 presents a summary of results and conclusions.

Throughout this paper we use a cosmology with
$H_0=70$ Km s$^{-1}$ Mpc$^{-1}$, $\Omega_m=0.3$, and $\Omega_\Lambda=0.7$.
All the magnitudes are in the Vega system, unless otherwise specified.

\section{Spectroscopic observations and data reduction}
\label{sec:observations}
The seven spheroidal galaxies belong to a sample of 23 spheroids candidate
cluster members,
photometrically selected in the field around the cluster XLSSJ0223-0436 at $z=1.22$
\citep{andreon05,bremer06}.
Briefly, the target galaxies were selected among all the galaxies brighter 
than $z_{850}<24$ 
within a projected radius D$\le1$ Mpc from the cluster centre.
Then, according to their $i_{775}-z_{850}$ colour, galaxies were selected 
within $\pm$0.2 mag from the peak of the color distribution centered
at the color of the brightest cluster members.
Finally, a visual classification on the F850LP images was used to select 
only those galaxies with elliptical/spheroidal morphology.
A detailed description of the whole sample, the structural parameters of galaxies
and their scaling relations are given in \cite{saracco17}.
\begin{table*}
\begin{minipage}[t]{1\textwidth}
\caption{List of observed galaxies in the XLSSJ0223 field and
spectroscopic redshift measurements. }
\label{tab:sample}
\centerline{
\begin{tabular}{rccccccccccr}
\hline
\hline
  ID &    RA      &        Dec   &  F850LP & $i_{775}-z_{850}$& log$\mathcal{M}_*$& log$\mathcal{M}_{dyn}$& $z_{spec}$ & Em$^a$& F([OII])$^b$ & SFR& S/N$^c$\\
     &   (h:m:s)& (d:p:s)&   (mag) & (mag) &(M$_\odot$) &(M$_\odot$) & & & (erg cm$^{-2}$ s$^{-1}$)& (M$_\odot$ yr$^{-1}$)&\\
\hline
 651&    02:23:05.759&  -04:36:10.27&  21.62$\pm$0.01  & 1.09$\pm$0.02 &  10.94&11.29& 1.2192&  [NeV]& 6$\pm$2& 0.7$\pm0.2$  	     & 11.3   \\ 
 972&    02:23:04.718&  -04:36:13.47&  22.61$\pm$0.03  & 0.92$\pm$0.03 &  10.54&11.32& 1.2153&  ...&   $<$6      &$<$0.8   	      &  6.2  \\ 
1142&    02:23:03.262&  -04:36:14.60&  21.30$\pm$0.01  & 1.00$\pm$0.02 &  11.50&11.68& 1.2204&  ...&  $<$1.5     &$<$0.2   	      & 12.7  \\  
1370&    02:23:02.021&  -04:36:43.26&  23.10$\pm$0.03  & 0.96$\pm$0.04 &  10.08&10.10& 1.2249&  ...&  $<$7      &$<$0.9   	      &  5.6  \\  
1442&    02:22:57.980 & -04:36:22.31&  21.88$\pm$0.01  & 0.91$\pm$0.02 &  10.80&10.67& 1.2250& [OII] & 27$\pm$4 & 3.0$\pm0.4$      & 11.3  \\  
1630&    02:23:00.929&  -04:36:50.19&  21.48$\pm$0.01  & 1.04$\pm$0.02 &  10.75&11.18& 1.2109&  [OII]& 34$\pm$6 & 4.0$\pm0.6$      & 12.2   \\  
1711&    02:22:59.990&  -04:36:02.53&  20.92$\pm$0.01  & 1.02$\pm$0.02 &  11.20&11.52& 1.2097&        ... & ... &...  		      & 18.2 \\  
\hline
\end{tabular}
}
{$^a$ Emission lines detected: dots - no emission line $>$1$\sigma$.\\
$^b$ Flux is in units of $10^{-18}$ erg cm$^{-2}$ s$^{-1}$.\\
$^c$ S/N per \AA ngstrom in the rest-frame of the galaxy, estimated in the
interval $\sim$4000-4150 \AA.}
\end{minipage}
\end{table*}

Spectroscopic observations of the targets 
were performed in multi-object spectroscopic
mode (MOS) with the Multi-Object Double Spectrograph (MODS, 1 and 2) 
\citep{pogge10} mounted at the Large Binocular Telescope (LBT).
Observations confirmed 13 spheroidal galaxies
members of the cluster XLSSJ0223.
The whole specroscopic data will be presented in a forthcoming paper
(Saracco et al. 2018, in preparation).
Here, we focus the analysis on the seven spheroids having the highest S/N,
suited to stellar population study.

Observations were carried out 
with filter GG495 coupled with the grim G670L sampling the wavelength range 
0.5$\mu$m$<\lambda<1.0\mu$m  at 0.85 \AA/pix.
We adopted a slit width of 1.2'' resulting in a spectral resolution 
R$\simeq 1150$, corresponding to a FWHM$\simeq$7.4 \AA\ at 8500\AA.
A bright star was put in a slit to accurately
measure the offsets in the observing sequence and perform the
correction for telluric absorption lines (see below).
Observations, consisting in a sequence of exposures (ABBA) 
of 900 sec each at dithered positions offset by $\sim5$",
were collected both with MODS1 and MODS2
for a total effective integration time of 8 hours.

Raw data were first pre-processed by the MODS pipeline that applies 
correction for bias-subtraction, flat-field (pixel-to-pixel) variation 
and for optical distortions, providing also the wavelength calibration 
through the inverse dispersion solution (0.08\AA\ rms accuracy).
Then, standard reduction was performed with IRAF tasks.
A first sky-subtraction was applied by subtracting from each frame
the following one in the dithering observing sequence.
The sky-subtracted frames were aligned to sub-pixel scale 
using {\it drizzle} resampling algorithm, and than co-added.
The shifts to align the images were estimated using the 
positions of the peak of the spectrum collapsed along the wavelengths 
of the bright star present in the masks.
Before co-adding, MODS1 and MODS2 
2D spectra were corrected for the relative spectral slit flats, 
obtained using quarz-halogen flat fields taken through the MOS masks,  
and for the relative sensitivity function.
The sensitivity function was derived from the spectrum of the same 
spectrophotometric standard star obtained with both cameras.

The 1D spectrum of each galaxy was extracted using the Iraf task $apall$.
After extraction, an additional residual sky-subtraction was
applied to remove  
sky residuals 
due to sky intensity variations between subsequent images in the ABBA 
sequence. 
To this end, we subtracted from the 1D object spectrum, 
the 1D spectrum of sky residuals obtained by averaging the pixels 
above and below the object, along the spatial direction.
Finally, we applied the software MOLECFIT 
\citep{kausch15} to perform telluric absorption correction, by fitting 
spectral regions of the bright star spectrum with prominent telluric
absorption. 
The resulting telluric absorption  model provided by MOLECFIT 
was applied to correct all galaxy spectra.

\section{Measurement of redshift and absorption line indices}
\subsection{Redshift measurement}
Measurements of spectroscopic redshifts and velocity dispersion were 
performed using stellar absorption features, 
by fitting the observed spectra with 
MILES simple stellar population (SSP) models \citep{vazdekis10}.
These models are primarily based on the MILES \citep{sanchez06,falcon11} 
and 
Indo-U.S \citep{valdes04} stellar libraries and have a spectral 
resolution of 2.5\AA\ 
\cite[][]{beifiori11}, close to the rest-frame resolution
($\sim$3.3\AA) of our spectra.
Spectral fitting was performed using the penalized
PiXel-Fitting method \cite[pPXF,][]{cappellari04,cappellari17}.

\begin{figure*}
	\includegraphics[width=11.5truecm]{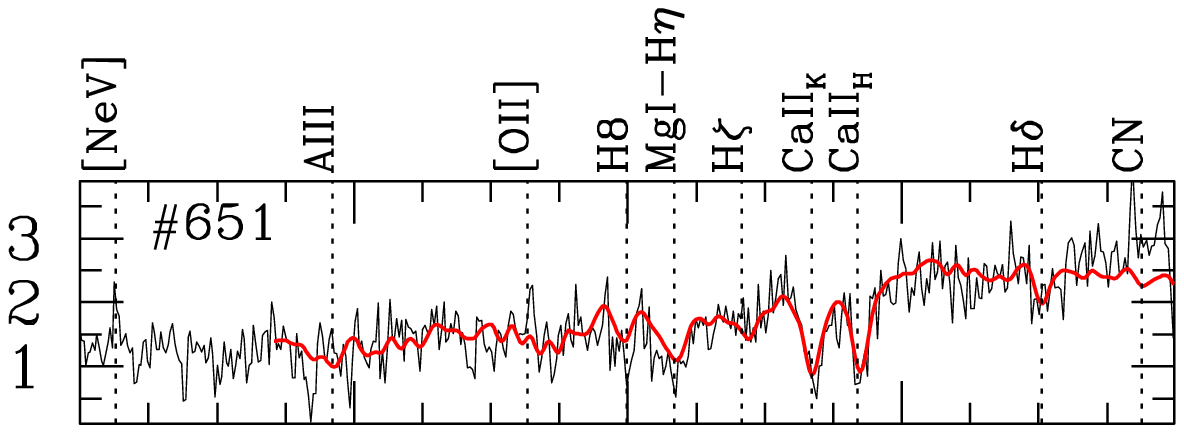}
	\includegraphics[width=2.2truecm]{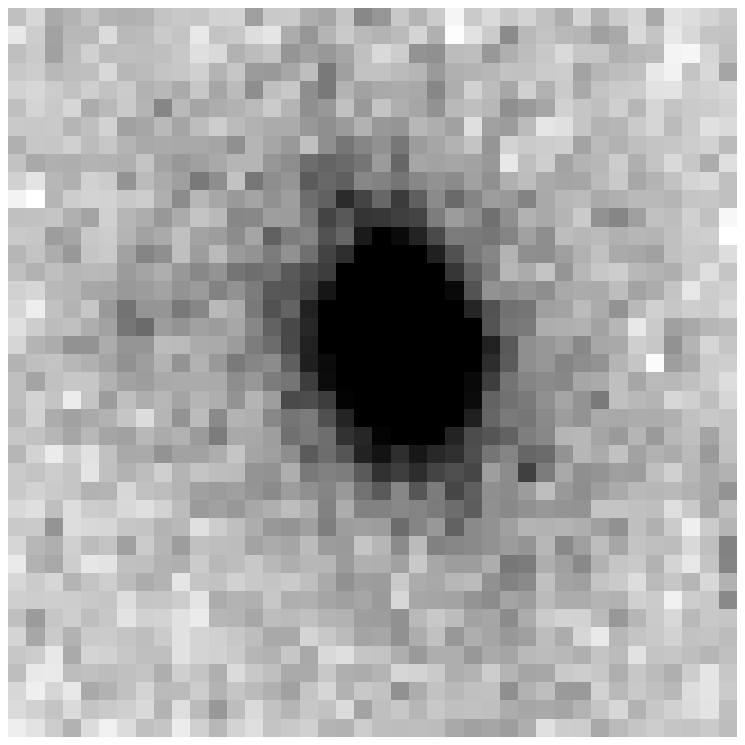}
	\includegraphics[width=11.5truecm]{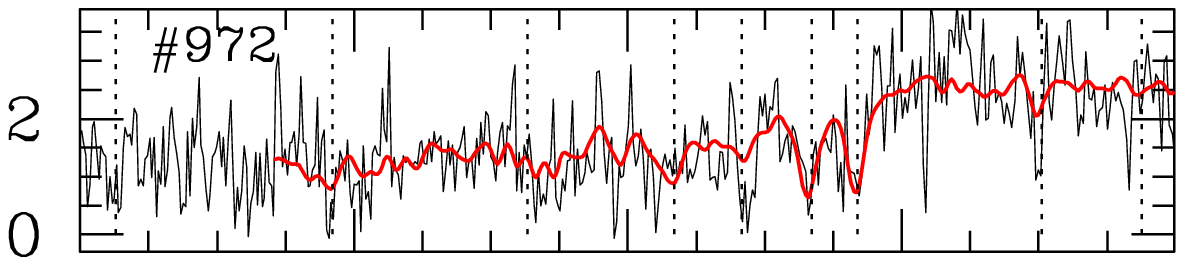}
	\includegraphics[width=2.2truecm]{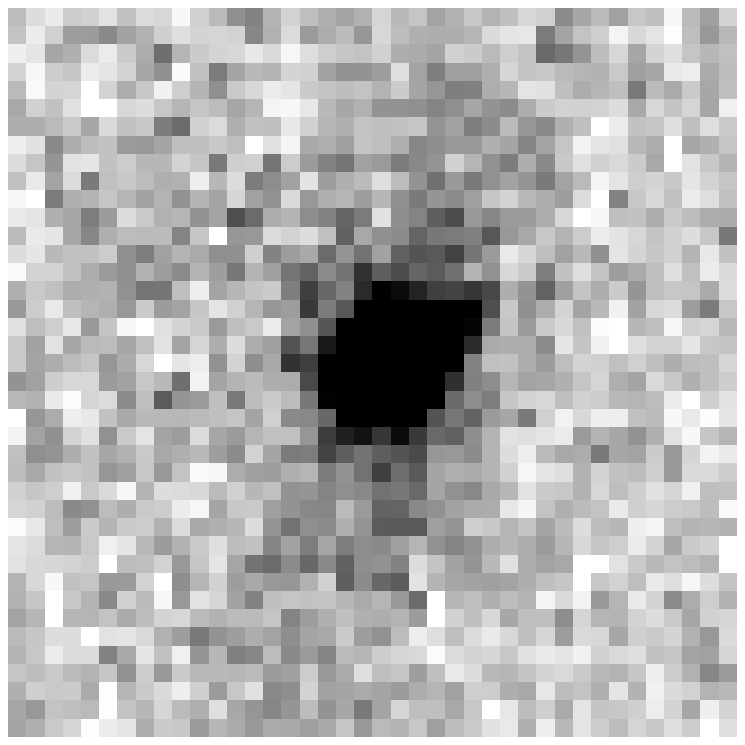}
	\includegraphics[width=11.5truecm]{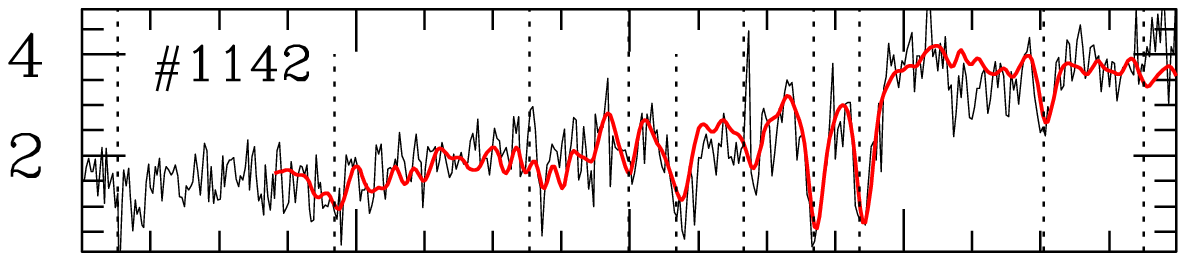}
	\includegraphics[width=2.2truecm]{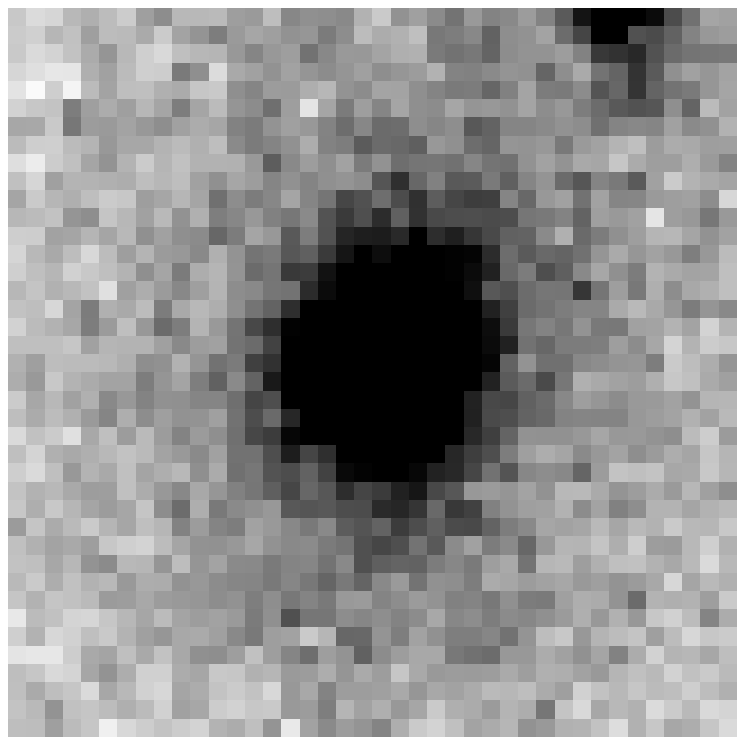}
	\includegraphics[width=11.5truecm]{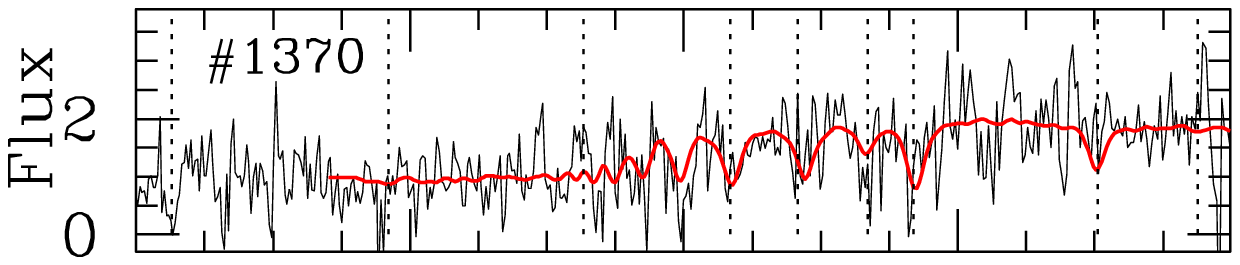}
	\includegraphics[width=2.2truecm]{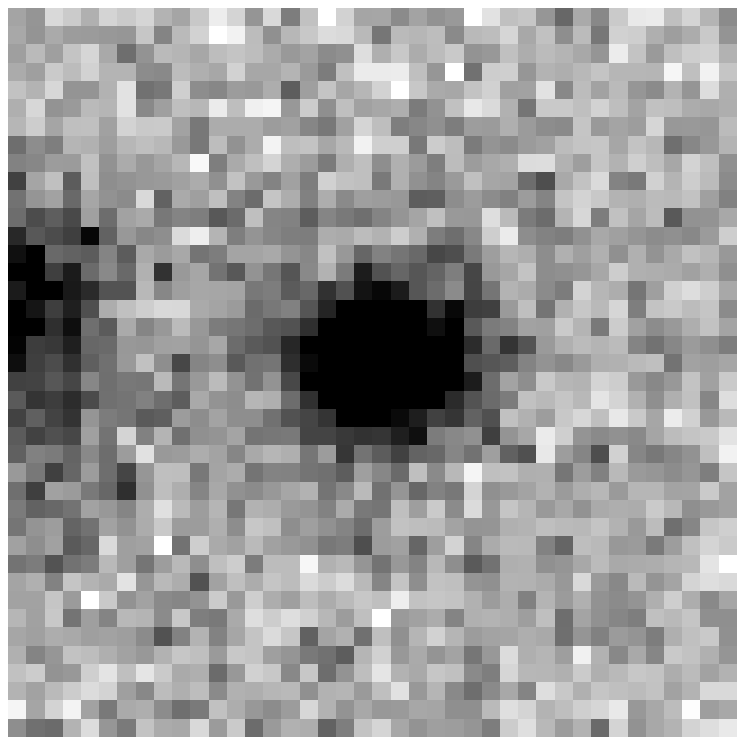}
	\includegraphics[width=11.5truecm]{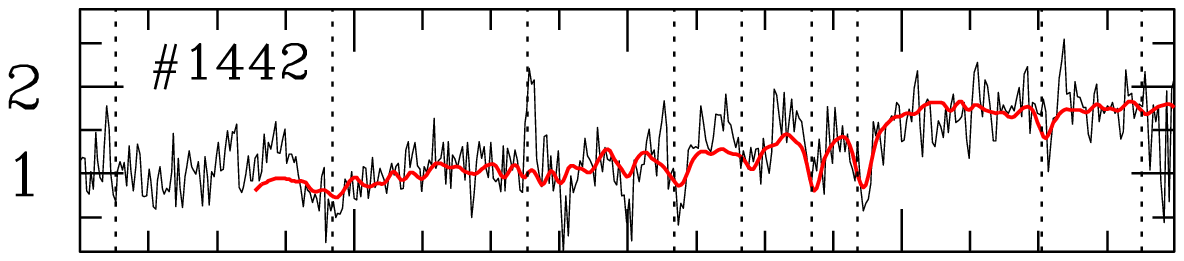}
	\includegraphics[width=2.2truecm]{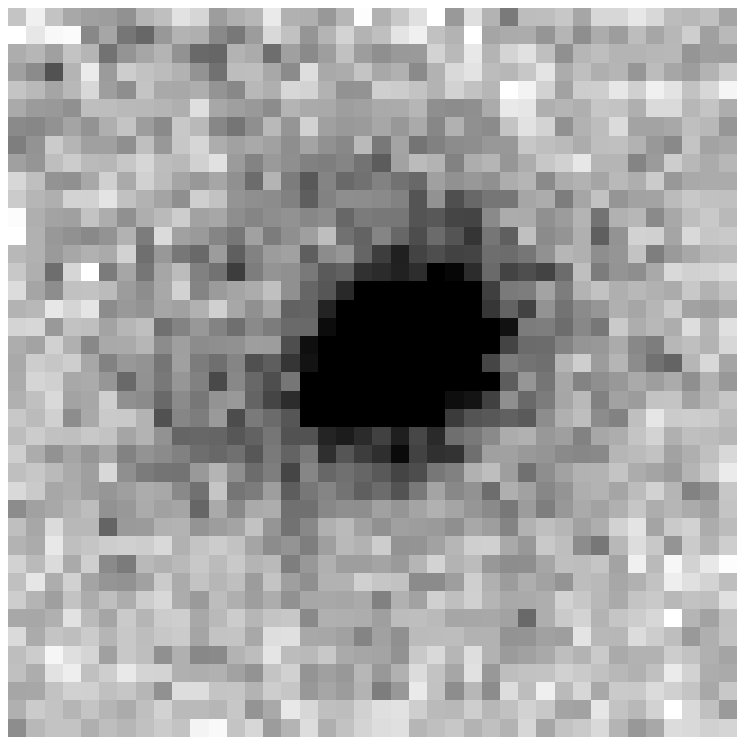}
	\includegraphics[width=11.5truecm]{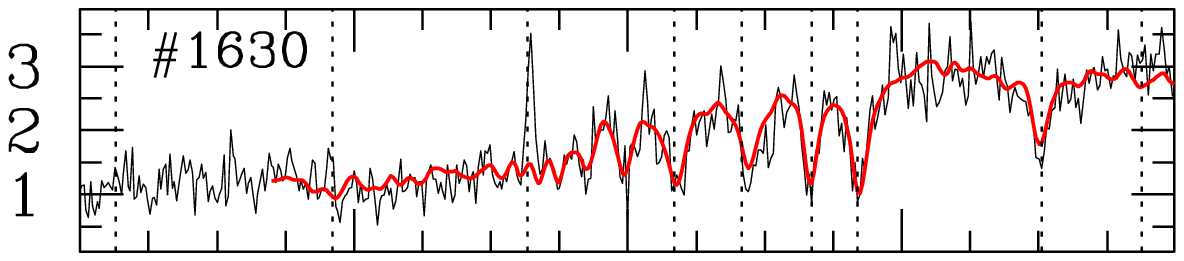}
	\includegraphics[width=2.2truecm]{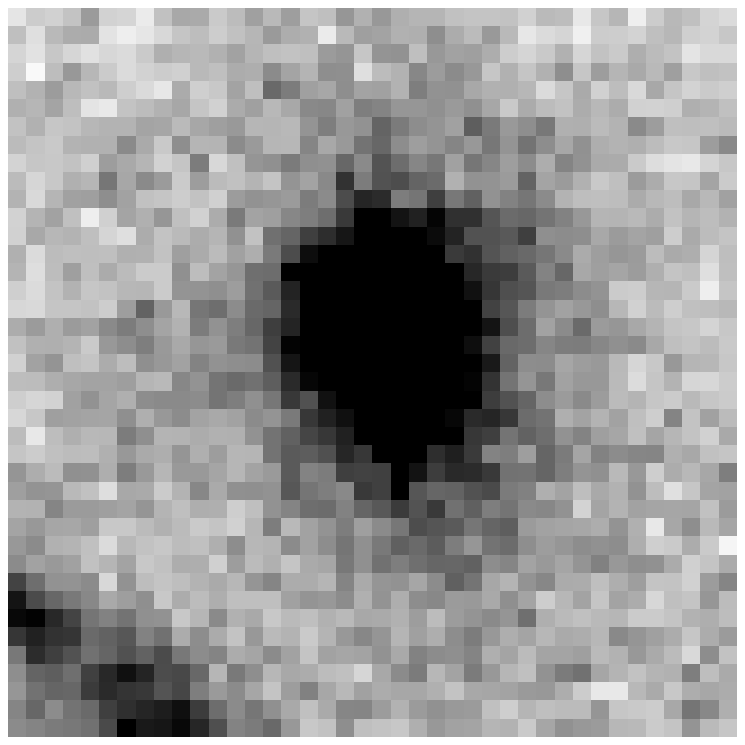}
	\includegraphics[width=11.5truecm]{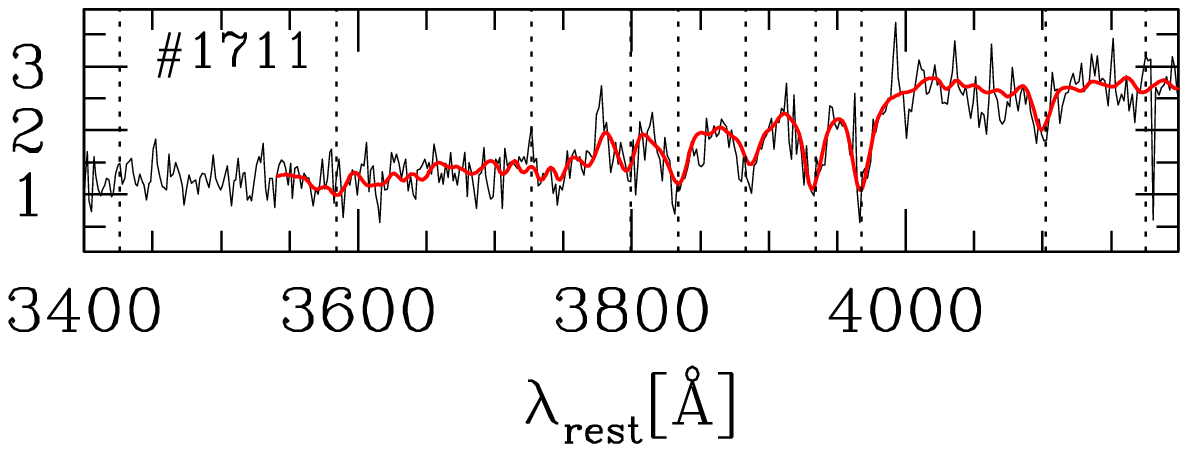}
  	\includegraphics[width=2.2truecm]{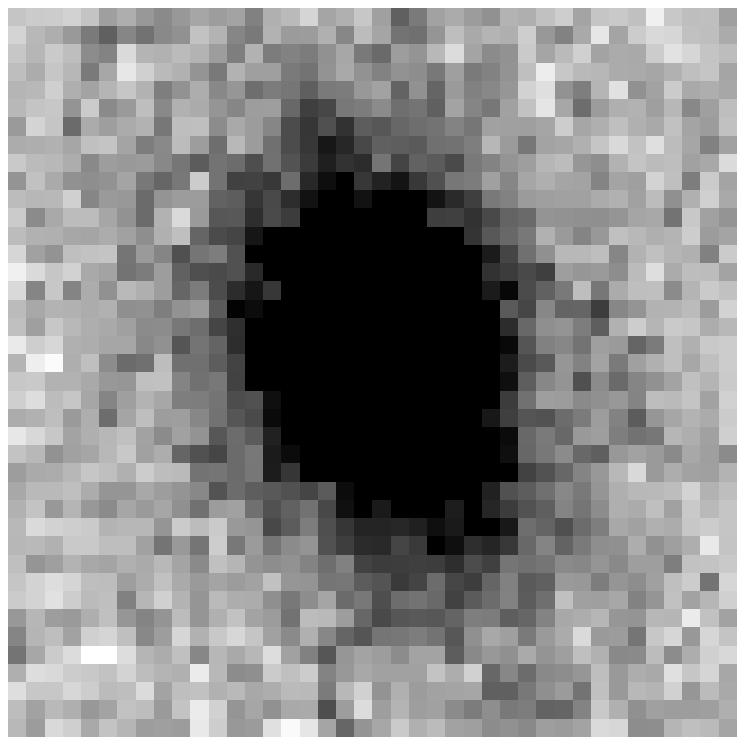}
   \caption{\label{fig:7spectra} MODS spectra of the seven cluster members spheroidal galaxies for which we
   derived the velocity dispersion.   The black curve is the observed spectrum 
    arbitrarily normalized, binned to 3.4 \AA/pixel. 
   These spectra have S/N=[5-18] per Angstrom in the restframe of the galaxy, in the range 
   4000$<\lambda_{rest}<$4150\AA. 
   The red curve is the best-fitting MILES model resulting from the pPXF spectral fitting. 
   The dotted vertical lines mark the main spectral features labelled on top
   of the figure. To the right of each spectrum, the ACS-F850LP image
   (2$\times$2 arcsec) of the target galaxy is shown.}
\end{figure*}

Velocity dispersion measurements for the whole set of spectroscopic
observations 
 will be presented in a forthcoming paper (Saracco et al. 2018, in preparation).
{  Here, we summarize the method used.
For each spectrum, we derived the broadening of its absorption lines,
 $\sigma_{obs}$, by performing spectral fitting in the range 3550--4200 \AA. 
The galaxy velocity dispersion, $\sigma_\star$ is then derived by correcting the 
$\sigma_{obs}$ for the instrumental broadening of $\sim$111 km/s, resulting from the 
instrumental resolution R$\simeq 1150$ (see \S 2).

The uncertanties on $\sigma_*$ have been derived 
by repeating the measurements on 100 simulated spectra obtained by
summing to the best-fitting template the 1D sky residuals extracted 
from the real 2D spectra, randomly shuffled in wavelength.
The typical uncertanties are in the range 10\%-25\%, with the exception of 
galaxy \#1370 affected by a much larger uncertaninty because of the low
S/N (see Tab. \ref{tab:sample}).
We have tested the robustness of the measurements against the library of template used 
and the wavelength range considered by repeating the fitting with a set of synthetic 
stars extracted from the library of \cite{munari05} 
and a set of SSPs of \cite{bruzual03},
and by slightly varying the fitted spectral range.
In all the cases, we obtained measurements well within the estimated errors. 
}
 
The seven  spheroids cluster members here analysed have  S/N$\sim$5-18 per \AA\ 
in the rest-frame interval 4000-4150\AA, that turned out to be sufficient 
to perform the stellar population analysis, as detailed below.
In Tab. \ref{tab:sample} we report the main properties of the seven spheroidal 
galaxies while their spectra are shown in Fig. \ref{fig:7spectra}.
In each panel, the black curve plots the 8-hours MODS spectrum binned
to 3.4 \AA/pix and the red curve is the best-fitting pPXF template.
For each galaxy we also show the $2\times2$ arcsec ACS-F850LP image.

\subsection{Emission lines: AGN and star formation}
Two galaxies, namely \#1442 and \#1630 (see Tab. \ref{tab:sample}), 
show clearly [OII] line emission, while
galaxy \#651 shows a weak [OII] emission accompanied by the presence of [NeV] 
line emission.
The high-ionization [NeV]$\lambda3426$ emission line is considered a signature
 of nuclear activity \citep{schmidt98,gilli10} since stars do not reach such high 
ionization potential \cite[e.g.][]{haehnelt01}, therefore galaxy \#651 most probably 
hosts an AGN.
Apart from \#651, there is no evidence for the presence of AGN
in any other galaxy.

Assuming that [OII] emission is due to star formation (also for \#651),  we derived 
the star formation rate (SFR) from the relation 
\begin{equation}
 SFR=1.4\cdot 10^{-41}L([OII])
\end{equation}
where the SFR is expressed in M$_\odot~yr^{-1}$ and the luminosity L([OII])
in erg$~s^{-1}$ \citep{kennicutt98}.
The luminosity L([OII]) has been derived from the line flux F([OII]) measured on the 
calibrated spectra, $L([OII])=4\pi d^2_L\cdot F([OII])$, where $d_L=8449.9$ Mpc is 
the luminosity distance of the cluster at $z=1.22$ for the adopted cosmology.
The measured [OII] fluxes and the resulting SFR are reported in Tab. \ref{tab:sample}
for those galaxies for which we detect an [OII] line flux $>$1$\sigma$.
Our spectra show that at least two spheroids out of the seven show evidence of 
weak star formation, at a rate $<$4 M$_\odot~yr^{-1}$, and one hosts an AGN.

We notice that, early-type galaxies with similar values of SFR 
(few M$_\odot~yr^{-1}$ or lower), are not rare and are observed 
both in the field and in cluster.
SFRs of about 2-3 M$_\odot~yr^{-1}$ are observed, for instance, in some early-types  
in cluster RDCSJ0848 at $z\sim1.27$ \citep[e.g.][]{jorgensen14} as well
as in the field at similar redshift \cite[e.g.][]{cimatti08,belli17}.
Also at lower redshift, some spheroidal galaxies exhibit low levels
of star formation, both in clusters \citep[e.g.][]{jorgensen17} and
in the field \cite[e.g.][]{fukugita04,huang09,yates14,george17}.
This star formation activity in field early-types is thought to be fueled by 
inflowing gas and/or gas-rich minor mergers \citep{belli17,george17}.
As to our seven spheroids, it is unlikely that star formation is fueled
by inflowing gas, since gas and galaxies should be in thermal
equilibrium in a cluster.
Moreover, minor mergers are disfavoured in the
cluster center regions given the large relative velocities of galaxies
\citep[e.g.][]{harrison11,treu03}. 
Thus, the low levels of star formation we observe could be the sign
of a more complex or longer star formation history or the quenching
phase of the main burst recently occurred.
We furthur discuss the presence of star formation below.

\subsection{Absorption line indices}
We measured absorption line Lick indices, following the definition by \cite{worthey97}
and \cite{trager98}, and the indices by \cite{rose94} in the rest-frame wavelength range
 covered by the observations.
To perform the measurements we made use of the software 
\texttt{LECTOR}\footnote{http://www.iac.es/galeria/vazdekis/vazdekis\_software.html}.
Besides these absorption indices, we measured the strength of the 4000\AA\ break
according to the D4000 definition by \cite{bruzual83} \citep[see also][]{gorgas99}
and the D$_n$ index by \cite{balogh99}.
The uncertainties on the spectral indices have been derived through simulations 
according to the following procedure.
For each observed 1D spectrum, we constructed a set of 100 simulated spectra
having the same S/N of the real one.
The S/N was evaluated in the rest-frame wavelength range 4000-4150 \AA.
The simulated spectra were obtained by summing to the best-fitting 
model template the 1D sky residual extracted from the final 2D spectra.
In each simulated spectrum, the values of the residuals  within the 
rest-frame wavelength range 3700-4400 \AA\ were randomly shuffled.
Then, for each simulated spectrum, we measured the indices using the same procedure
used for the real spectra.
We adopted as uncertainty on the measured index the median absolute deviation 
(MAD) resulting from the distribution of the 100 measurements.

The measured indices were corrected for galaxy velocity dispersion.
The corrections were obtained for each galaxy by comparing the
indices measured on the best-fitting model (smoothed to the $\sigma$ of the galaxy),
and those of the same model at the nominal resolution of the MILES
spectral library. 
The corrected measured Lick indices and Rose's indices with their errors are 
summarized in Tab. \ref{tab:lick} and Tab. \ref{tab:rose} respectively.
 
\begin{table*}
\begin{minipage}[t]{1\textwidth}
\caption{\label{tab:lick} Lick spectral indices.}
\centerline{
\begin{tabular}{rrcccrrrrrr}
\hline
\hline
  ID & CN3883 & CaHK&  D4000 & D$_{n}$  &  H$\delta_A$ & H$\delta_F$&CN1  &CN2 &
  Ca4227& G4300   \\
     &(mag)  &(\AA) &  	& &(\AA)  & (\AA) & (mag)  &(mag)   & (\AA)  &(\AA) \\
\hline
 651&    0.27$\pm$0.05 & 23.6$\pm$2.3   &1.86$\pm$0.05  & 1.51 &  3.1$\pm$2.0 & 3.7$\pm$1.0  & -0.21$\pm$0.05 & -0.20$\pm$0.06 & 4.2 $\pm$0.8  & 4.9$\pm$1.4 \\  
 972&    0.32$\pm$0.05 & 22.4$\pm$2.6   &2.13$\pm$0.05  & 2.00 &  1.3$\pm$2.2 & 4.6$\pm$1.2  &  0.17$\pm$0.05 &  0.25$\pm$0.07 & 3.9 $\pm$0.9  & 1.2$\pm$1.6 \\  
1142&    0.20$\pm$0.04 & 30.6$\pm$2.0   &2.19$\pm$0.04  & 1.75 &  0.7$\pm$1.7 & 4.1$\pm$0.9  &  0.03$\pm$0.04 &  0.07$\pm$0.05 & 0.2 $\pm$0.7  &10.4$\pm$1.2 \\  
1370&    0.04$\pm$0.06 & 19.8$\pm$3.0   &1.52$\pm$0.06  & 1.30 & -2.8$\pm$2.5 & 0.2$\pm$1.3  &  0.09$\pm$0.06 &  0.13$\pm$0.08 & 3.3 $\pm$1.0  & 6.2$\pm$1.8 \\  
1442&    0.15$\pm$0.05 & 17.7$\pm$2.3   &1.77$\pm$0.05  & 1.33 & -2.3$\pm$2.0 & 2.0$\pm$1.0  & -0.17$\pm$0.05 &  0.17$\pm$0.06 &-0.9 $\pm$0.8  &12.6$\pm$1.4 \\  
1630&   -0.01$\pm$0.04 & 10.0$\pm$2.0   &1.82$\pm$0.04  & 1.48 &  7.9$\pm$1.7 & 5.9$\pm$0.9  & -0.19$\pm$0.04 & -0.15$\pm$0.05 &-1.6 $\pm$0.7 & 2.7$\pm$1.2 \\  
1711&    0.15$\pm$0.03 & 18.5$\pm$1.8   &1.84$\pm$0.03  & 1.54 &  3.1$\pm$1.5 & 3.1$\pm$0.8  &  0.01$\pm$0.03 &  0.01$\pm$0.04 & 3.3 $\pm$0.6	& 6.8$\pm$1.0 \\  
\hline
\end{tabular}
}
{The indices are corrected for galaxy velocity dispersion (see Sec. 3.3).}
\end{minipage}
\end{table*}

\begin{table*}
\begin{minipage}[t]{1\textwidth}
\caption{\label{tab:rose} Same as Tab. \ref{tab:lick} but for Rose indices.}
\centerline{
\begin{tabular}{rcccccccc}
\hline
\hline
  ID &  3888/3859& CaII(H/K)& H$\delta$/FeI[4045] &H$\delta$/FeI[4063]  &p[Fe/H]  \\
     &   & & 	& &               \\
\hline
 651&  1.1$\pm$0.4&  0.8$\pm$0.4  & 1.1$\pm$0.2 & 1.1$\pm$0.2 & 1.0$\pm$0.2      \\  
 972&  0.3$\pm$0.4&  0.9$\pm$0.4  & 0.3$\pm$0.2 & 0.4$\pm$0.2 & 1.4$\pm$0.2      \\  
1142&  0.8$\pm$0.3&  1.2$\pm$0.3  & 0.9$\pm$0.2 & 0.9$\pm$0.2 & 1.3$\pm$0.1      \\  
1370&  0.7$\pm$0.5&  0.9$\pm$0.5  & 0.8$\pm$0.4 & 0.8$\pm$0.4 & 1.3$\pm$0.3      \\  
1442&  0.8$\pm$0.4&  0.7$\pm$0.4  & 1.1$\pm$0.2 & 1.1$\pm$0.2 & 1.0$\pm$0.2      \\  
1630&  0.5$\pm$0.3&  1.0$\pm$0.3  & 0.6$\pm$0.2 & 0.6$\pm$0.2 & 1.0$\pm$0.1      \\  
1711&  0.7$\pm$0.2&  0.9$\pm$0.2  & 0.7$\pm$0.1 & 0.7$\pm$0.1 & 1.1$\pm$0.1      \\  
\hline
\end{tabular}
}
\end{minipage}
\end{table*}

\section{Stellar population parameters}
The effective rest-frame wavelength range covered by our spectroscopic
observations (3500$<$$\lambda_{rest}$$<$4400\AA), does not provide us 
with enough spectral features suited to constrain the abundance ratios 
of $\alpha$-elements (Ca, Mg, Si, Ti etc.).
In particular, prominent tracers of Magnesium abundance, such
as the Mgb spectral feature at $\lambda \sim 5270 \, \AA$, usually considered the best
tracer of $\alpha$-elements, fall outside our spectral range.
For this reason, in the whole analysis, we limit ourselves to constrain the total 
metallicity [Z/H] without considering the effect of [$\alpha$/Fe].

\label{sec:lines}
\begin{figure*}
\includegraphics[width=16truecm]{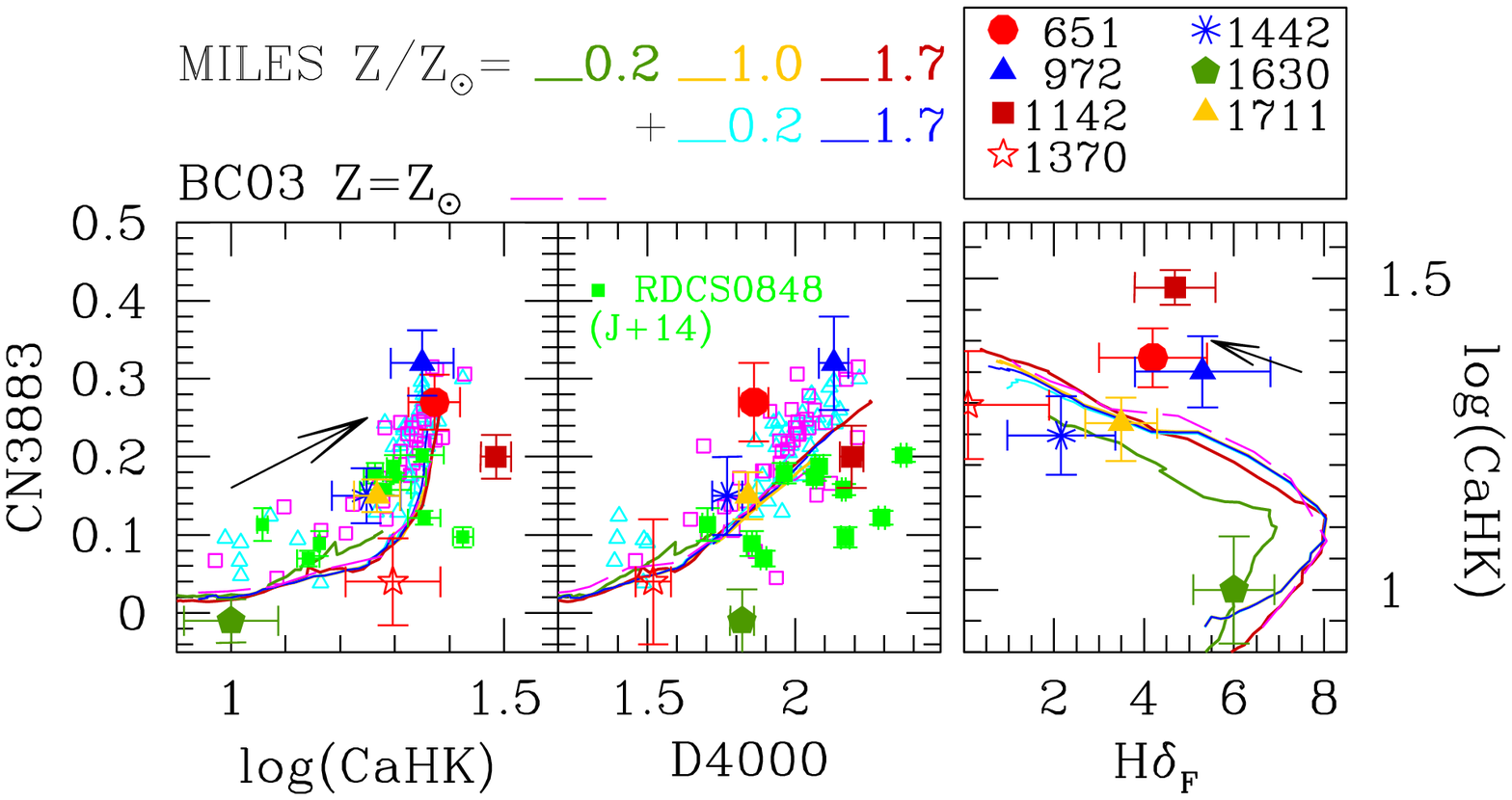}
\vskip -0.7truecm
\includegraphics[width=16truecm]{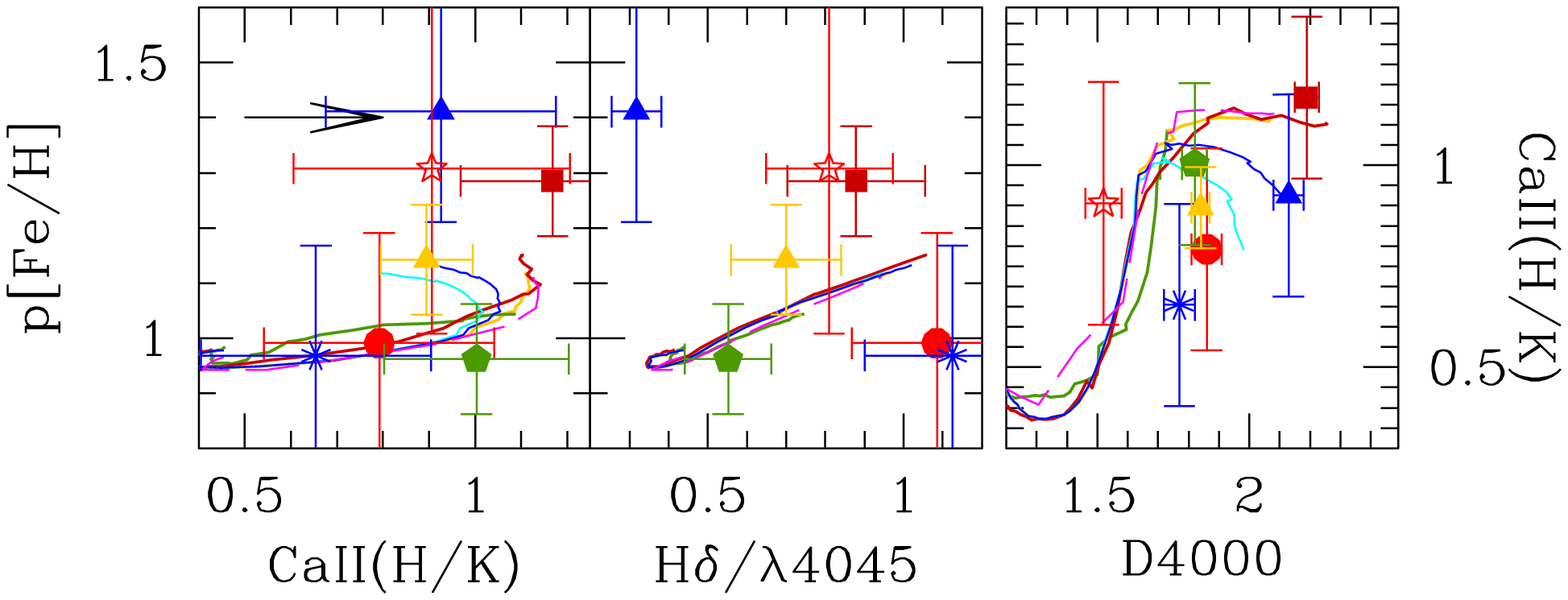}
\caption{\label{fig:lick} Absorption line strengths versus each other. 
Upper panels - Absorption Lick indices. 
The seven spheroidal galaxies belonging to the cluster 
XLSSJ0223 whose measurements are reported in Tab. \ref{tab:lick}, are represented
by the different symbols reported in the inset.
Green filled square are measurements of galaxies in the cluster RDCS0848 at $z\sim1.27$  
\citep{jorgensen14}.
Open (cyan) triangles and (magenta) squares are galaxies in the clusters 
MS0451.6 at $z\simeq0.54$  and RXJ1226.9 at $z\simeq0.89$ respectively \citep{jorgensen13}.
Solid curves are predictionis from MILES SSP models \citep{vazdekis10} for the three different 
metallicity values 
reported in the figure legend and plotted in the age interval 0.1-5 Gyr.
The cyan and blue thin curves are the Z$_\odot$ SSP in the interval 0.1-5 Gyr with the addition 
of a young (0.06 Gyr) component (0.5\% of the mass) with Z=0.2Z$_\odot$ and Z=1.7Z$_\odot$,
respectively.
The dashed (fuchsia) curve show predictions from BC03 SSP model \citep{bruzual03} with 
Chabrier IMF and solar metallicity in the same age interval.
The arrows indicates the direction in which the age increases following the models.
Observed indices are corrected for velocity dispersion.
Lower panels - Rose's \citep{rose94,rose94b} absorption line indices. 
The values are reported in Tab. \ref{tab:rose}.
Symbols are as in the upper panels.  
}  
\end{figure*}
\subsection{Constraints from absorption line indices}
In Fig. \ref{fig:lick} the main Lick absorption indices (upper panel) and the main
Rose's indices (lower panel) are plotted versus each other.
The upper panels show the strength of CN3883 versus CaHK and D4000, and the strength 
of CaHK versus H$\delta_F$.
The three plots are arranged to show two different metal lines, one
against the other (CN3883 vs CaHK), and against two different age sensitive indices
(D4000 and H$\delta_F$). 
The index CN3883 has been derived according to the definition of \cite{davidge94} 
\citep[see also][]{pickles85}.  
The lower panels follow the same criterion of the upper ones showing
the strength of p[Fe/H] versus CaII(H/K) and H$\delta$/Fe4045, and the strength of  
CaII(H/K) versus D4000.
The index p[Fe/H] is metal-abundance sensitive, H$\delta$/Fe4045 increases with
the stellar spectral type and the index CaII(H/K) is a good tracer of 
the presence of very young populations, being sensitive to  A-type stars 
\citep{rose94,longhetti99}. 

The seven spheroidal galaxies here analyzed are shown
with different (filled) symbols according to the legend in the inset.
Galaxies in cluster RDCS0848 (LinxW) at $z=1.27$, whose measurements 
have been obtained by \cite{jorgensen14} (green filled squares) are also shown.
For this cluster, we considered the galaxies belonging to samples \#5 and \#4 
of Tab. 7 in \cite{jorgensen14} for which measurements are available.
For comparison with lower redshift data, we show the measured indices for galaxies in 
cluster MS0451.6 at $z\simeq0.54$  (open cyan triangles) and in cluster RXJ1226.9 at 
$z\simeq0.89$ (open magenta squares) by \cite{jorgensen13}.

Solid curves are model predictions based on MILES SSPs \citep{vazdekis10} with Chabrier initial 
mass function (IMF) \citep{chabrier03}.
The models shown are plotted in the age range [0.1,5] Gyr and refer to three different values of 
metallicity [Z/H]: -0.71 (dark green curve), 0.0 (orange line) and +0.22 (dark red curve).
We also considered the two cases of a very young (0.06 Gyr) population representing 0.1\% of the
stellar mass with  [Z/H]=-0.71 (cyan curve) and [Z/H]=0.22 (blue curve) superimposed to 
an underlying dominant [Z/H]=0 population.  
Dashed (fuchsia) curve represents the prediction based on BC03 SSPs \citep{bruzual03} with solar 
metallicity and Chabrier IMF, in the same age range.
No significant differences arise among the different models considered.
Therefore, the results of the following analysis are independent of the adopted models. 

In spite of the degeneracy between age and metallicity affecting most of the indices,
the diagnostic diagrams in Fig. \ref{fig:lick}, can be used to constrain relative 
differences in age and metallicity among galaxies in our sample. 
The deviation from SSPs could suggest a more complex history of star formation than the 
one described by a SSP as found, for instance, by \cite{lonoce14} for field early-type galaxies 
at $\sim1$.
Whether this is the case, it will be explored in the next section through full spectral fitting.

Galaxies \#1370, \#1442 and \#1630 lie on or below the lowest metallicity track in the 
CaHK vs H$\delta_F$ diagram, the one less affected by the degeneracy, while galaxies 
\#651, \#972 and \#1142 lie above the highest metallicity track and galaxy \#1711 lies in between.
This behaviour is due to the fact that the strength of CN3883 in galaxies 
\#1370, \#1442, \#1630 and \#1711 is significantly 
weaker than in galaxies \#651, \#972 and \#1142, as visible also from the  CN3883 vs CaHK diagram.
Also the p[Fe/H] index tends to be weaker for galaxies \#1442, \#1630 and \#1711
with respect to the other galaxies in the sample.
These differences are consistent with galaxies \#1370, \#1442,  \#1630 and \#1711 having lower 
metallicity than the remaining three galaxies.

As to the age, we notice that the D4000 is stronger in galaxies \#651, \#972 and \#1142 
than in the other galaxies, in agreement with the CaHK index being stronger in 
these three galaxies, and with the CaII(H/K) close to unity for \#972 and \#1142.
The possible presence of weak star formation (see \S 3.2) superimposed to an older population,
could justify the lower value of this index in galaxy \#651.
These three galaxies tend to occupy the end at oldest ages of the SSP tracks in the diagrams 
showing metal-sensitive index vs age-sensitive index (e.g. CN3883/CaHK and CN3883/D4000), 
at odds with the other galaxies.
Thus, the picture is consistent with an age for \#651, \#972 and \#1142 older than for the
other galaxies.

Galaxies \#1630 and \#1442 are among the youngest galaxies in our sample. 
They both show star formation (see Tab. \ref{tab:sample}) but are characterized by 
different values of the indices, with the exception of the D4000, not significantly different.
Galaxy \#1630 occupies the lowest-age end of the SSP tracks in most diagrams.
It is characterized by the highest H$\delta$ indices and by the lowest CaHK.
The Rose's index CaII(H/K), whose CaII(H) line is sensitive to young stars, 
is close to one suggesting that the whole stellar population of this galaxy is genuinely young. 
For comparison, galaxy \#1442 has a higher CaHK and lower H$\delta$ indices,
accompanied by a CaII(H/K) lower than unity. 
This would suggest that the newly formed stars are superimposed to an older population. 


\subsection{Absorption lines fitting}
As a first  step to characterize stellar population  properties of the
seven spheroidal galaxies in XLSSJ0223,  we have derived their age and
metallicity, [Z/H],  by  comparing  observed line-strengths  to
predictions from SSP models with varying age and metallicity.  
To this effect, we use the same set of MILES SSP models \citep{vazdekis10}
as for  spectral fitting (see  Sec. 5.1), namely SSPs based  on PADOVA
isochrones (Girardi  et al. 2000),  with a  Chabrier IMF, ages  in the
range   from  0.06   to  4.5~Gyr,   and  metallicity   in  the   range
$-2.32$--$0.22 \, Z_{\odot}$.

The fitting is performed by minimizing the expression:

\begin{equation}
 \chi^2 (age,[Z/H]) = \sum_j \frac{(O_j-M_j)^2}{s_j^2}
\end{equation}

where the index $\rm j$ runs over the selected set of Lick spectral
indices in the rest-frame range 3500--4400 \AA\ (see Tab. \ref{tab:lick}), 
$\rm O_j$ and $\rm M_j$ are observed and model index values (the latter
depending on age and $\rm [Z/H]$), and $\rm s_j$ are uncertainties on
observed indices.  
The resulting best-fitting age and metallicity are
reported in Tab. \ref{tab:parameters}.  
The quoted errors are obtained by running the
fitting procedure on a set of simulated spectra, having the same S/N
as the observed spectra.

It  is worth  noting  that, as  expected, the  age and  metallicity
values  in  Tab. \ref{tab:parameters}  confirm  the  general  trends  
obtained  from  the diagnostic diagrams in Fig. \ref{fig:lick} (see Sec.~4.1), 
namely that galaxy \#1630
is the youngest  object in our sample, galaxies \#651,  \#972, and \#1142
are the  oldest, while  galaxies \#1370  and \#1442  tend to  have lower
metallicities than the remaining systems.

\begin{table}
\caption{\label{tab:parameters} Age and metallicity estimates resulting from absorption lines fitting}
\centerline{
\begin{tabular}{rrr}
\hline
\hline
  ID &  Age&  [Z/H]  \\
     &   (Gyr) &     \\
\hline
 651&   2.2$_{-0.4} ^{+0.8}$  &  -0.31$_{-0.17}^{+0.44}$  \\  
 972&   1.9$_{-0.7} ^{+1.3}$  &  +0.02$_{-0.51}^{+0.19}$  \\  
1142&   2.8$_{-0.6} ^{+1.6}$  &  +0.16$_{-0.31}^{+0.06}$  \\  
1370&   1.3$_{-0.3} ^{+1.0}$  &  -0.05$_{-0.57}^{+0.18}$  \\  
1442&   1.9$_{-0.8} ^{+1.5}$  &  -0.52$_{-0.77}^{+0.50}$  \\  
1630&   0.9$_{-0.1}^{+0.2}$   &  +0.01$_{-0.39}^{+0.20}$  \\  
1711&   1.5$_{-0.4}^{+0.7}$  &   -0.10$_{-0.52}^{+0.32}$  \\  
\hline
\end{tabular}
}
\end{table}

\section{Constraints on star formation history}

\begin{table*}
\caption{\label{tab:fitting} Age and metallicity estimates resulting from spectral fitting}
\centerline{
\begin{tabular}{rlrrrrrcc}
\hline
\hline
  ID &  Age$_{L}$&  [Z/H]$_{L}$&  Age$_{M_*}$&  [Z/H]$_{M_*}$ & A$_V$& $z_{f}$&n(SSPs)& {$\chi^2_\nu$}\\
     &   (Gyr) &     & (Gyr) &  &(mag)& & &\\
\hline
 651&   2.7$_{-0.7} ^{+1.0}$  & +0.09$^{+0.13}_{-0.42}$ & 3.0$_{-0.7}^{+0.8}$  & +0.09$^{+0.13}_{-0.42}$ &  0.0 & 3.4  & 4 &{ 0.89}\\  
 972&   3.4$_{-0.6} ^{+0.9}$  & +0.16$^{+0.06}_{-0.16}$ & 4.4$_{-0.5}^{+0.1}$  & +0.21$^{+0.01}_{-0.21}$ &  0.3  &$>$6.5& 2&{ 0.82}\\  
1142&   2.3$_{-0.2} ^{+1.2}$  & +0.21$^{+0.01}_{-0.21}$ & 2.4$_{-0.2}^{+1.4}$  & +0.22$^{+0.00}_{-0.22}$ &  0.8  & 2.6  & 2&{ 0.93}\\  
1370&   1.3$_{-0.2} ^{+1.8}$  & -0.27$^{+0.49}_{-0.12}$ & 1.9$_{-0.5}^{+0.8}$  & -0.10$^{+0.32}_{-0.61}$ &  1.0  & 2.0  & 2&{ 0.80}\\  
1442&   1.7$_{-0.5} ^{+0.7}$  & -0.18$^{+0.44}_{-0.21}$ & 2.7$_{-0.5}^{+0.2}$  & -0.01$^{+0.23}_{-0.38}$ &  0.9  & 2.9  & 4&{ 0.90}\\  
1630&   1.2$_{-0.3}^{+0.6}$   & -0.13$^{+0.39}_{-0.26}$ & 1.4$_{-0.2}^{+0.7}$  & -0.22$^{+0.44}_{-0.17}$ &  1.2  & 1.9  & 3&{ 0.88}\\  
1711&   1.7$_{-0.5}^{+0.2}$   & +0.05$^{+0.17}_{-0.05}$ & 2.1$_{-0.5}^{+0.2}$  & +0.18$^{+0.04}_{-0.18}$ &  0.7  & 2.4  & 4&{ 0.90}\\  
\hline
\end{tabular}
}
\end{table*}
\subsection{Full spectral fitting}
To recover the main properties of the stellar populations of the seven galaxies, 
the mean stellar age and metallicity, and to constrain their star formation
history (SFH), we also used full spectral fitting.
In this approach, one assumes that the observed spectrum is the 
superposition of few simple stellar populations (SSPs), 
with age Age$_i$ and metallicity Z$_i$, each one contributing 
with a different weight,
extracted from a larger base of models of different ages and metallicities. 
To perform the spectral synthesis and find the best-fitting linear
combination of SSPs, we used the code \texttt{STARLIGHT} 
\citep{cid05,cid07, mateus07, asari07}.

We adopted as reference base of models the SSPs from the MILES-based 
library (v 11.0) of \cite{vazdekis10},
covering the wavelength range 3540-7400 \AA\ at a spectral resolution of 2.5\AA.
We considered Chabrier IMF and two different sets of {\it Base} SSPs: 
the first one based on the PADOVA00 isochrones  \citep{girardi00} and the second 
one on the BaSTI isochrones \citep{pietrinferni04}.
For both these sets of models, the ages considered are $\sim$30 in the range [0.06; 4.5] Gyr 
while the metallicities are 7 for the PADOVA set in the range [-2.23; 0.22],  and
9 for the BaSTI set in the range [-2.27; 0.26].
We allowed for internal reddening in the \texttt{STARLIGHT} fits, by considering
both the Cardelli \citep[CCM;][]{cardelli89} and the Calzetti 
\citep[HZ5;][]{calzetti00} extinction laws in the fit. 
{  The fitting was perfomed in the wavelength range 3600-4400\AA.}

To assess the robustness of the derived stellar population properties 
with respect to the models adopted in the fitting, we also considered
 a base composed of BC03 SSPs 
\citep{bruzual03} with Chabrier IMF, 20 ages in the range [0.06; 4.5] Gyr 
and 5 metallicities in the range [-1.7; 0.4], and a base composed
of \cite{maraston11} MILES-based models (MS11 hereafter) with Chabrier 
IMF and including 20 ages in the range [0.06; 4.5] Gyr and 5 metallicities 
in the range [-2.3; 0.3].
The results of the fitting with these two sets of models are summarized
in Tables \ref{tab:bc03} and \ref{tab:m11}, respectively.

\subsection{Ages and metallicities derivation}
We defined the luminosity-weighted age $\langle$Age$\rangle_{L}$ and metallicity
[Z/H]$_{L}$ as
\begin{equation}
\begin{split}
&{\rm Age}_{L}=\sum_i x_i Age_i\\
&{\rm [Z/H]}_{L}=log\sum_i x_i Z_i/Z_\odot
\end{split}
\end{equation}
where $x_i$ is the fractional contribution of the $i$-th SSP
considered in the synthesis, while $Age_i$ and $Z_i$ are its age and metallicity.
Analogously, we defined the stellar mass-weighted age and metallicity as
\begin{equation}
\begin{split}
 &{\rm Age}_{M}=\sum_i m_i Age_i\\
&{\rm [Z/H]}_{M}=log\sum_i m_i Z_i/Z_\odot
\end{split}
\end{equation}
where $m_i$ is the fraction of stellar mass of the $i$-th SSP.

\begin{figure*}
\includegraphics[width=15truecm]{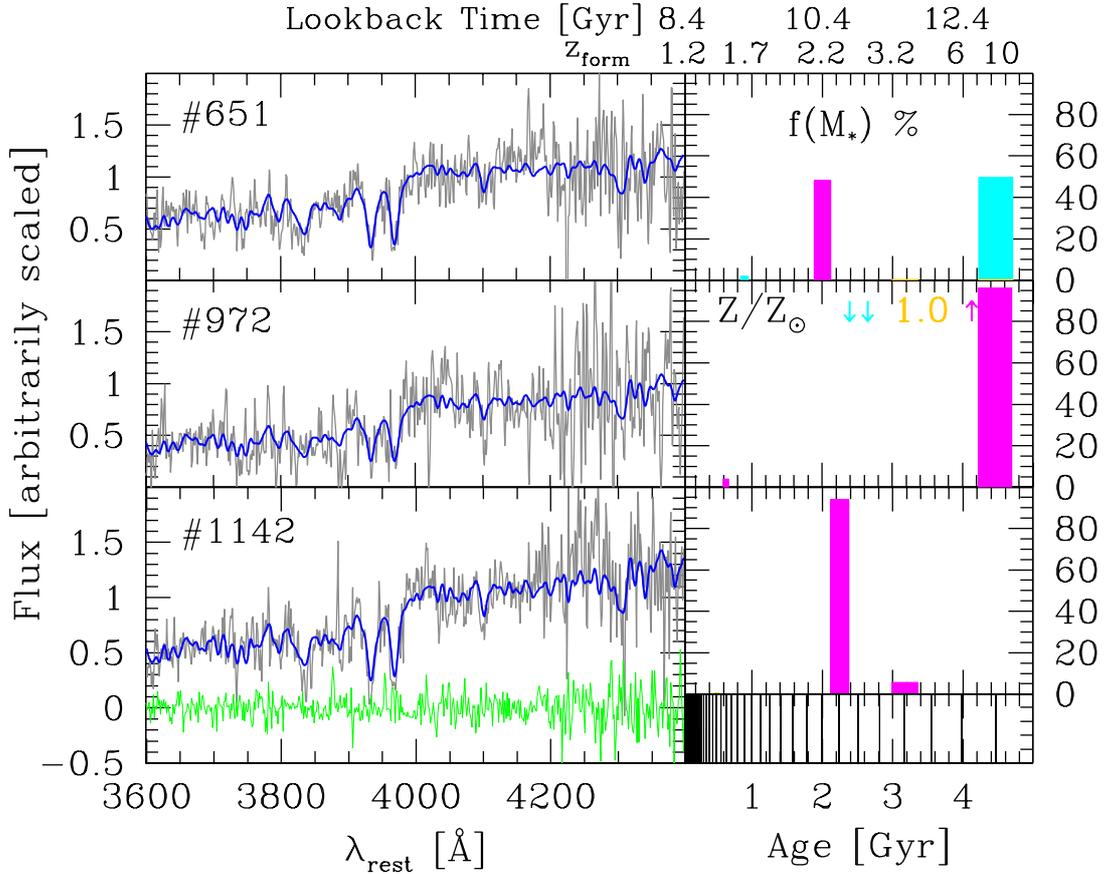}
\caption{\label{fig:star1} Results of the spectral synthesis for galaxies
\#651, \#972 and \#1142.
Left panels - The best-fitting composite spectrum (blue curve) resulting from
the spectral synthesis performed with \texttt{STARLIGHT} is superimposed
to the observed spectrum (dark gray curve) binned to 3.4\AA/pix.
The green curve in the bottom panel represents the typical sky residuals extracted from the image
of galaxy \#1142.     
Right panels - Star formation history of galaxies. Fraction of stellar mass of each 
SSP contributing to the best-fitting spectrum as a function of lookback time 
and formation redshift (top x-axis). 
Different colours encode different values of metallicity:
cyan Z$<$Z$_\odot$ (cyan down arrows in the legend), 
yellow  Z=Z$_\odot$, magenta Z$>$Z$_\odot$ (magenta upward arrows in the legend).
The bar code on the bottom represents the grid of ages considered
in the base of SSPs adopted to run \texttt{STARLIGHT} (see the text).
}  
\end{figure*}
\renewcommand{\thefigure}{\arabic{figure} (Cont.)}
\addtocounter{figure}{-1}

\begin{figure*}
\includegraphics[width=15truecm]{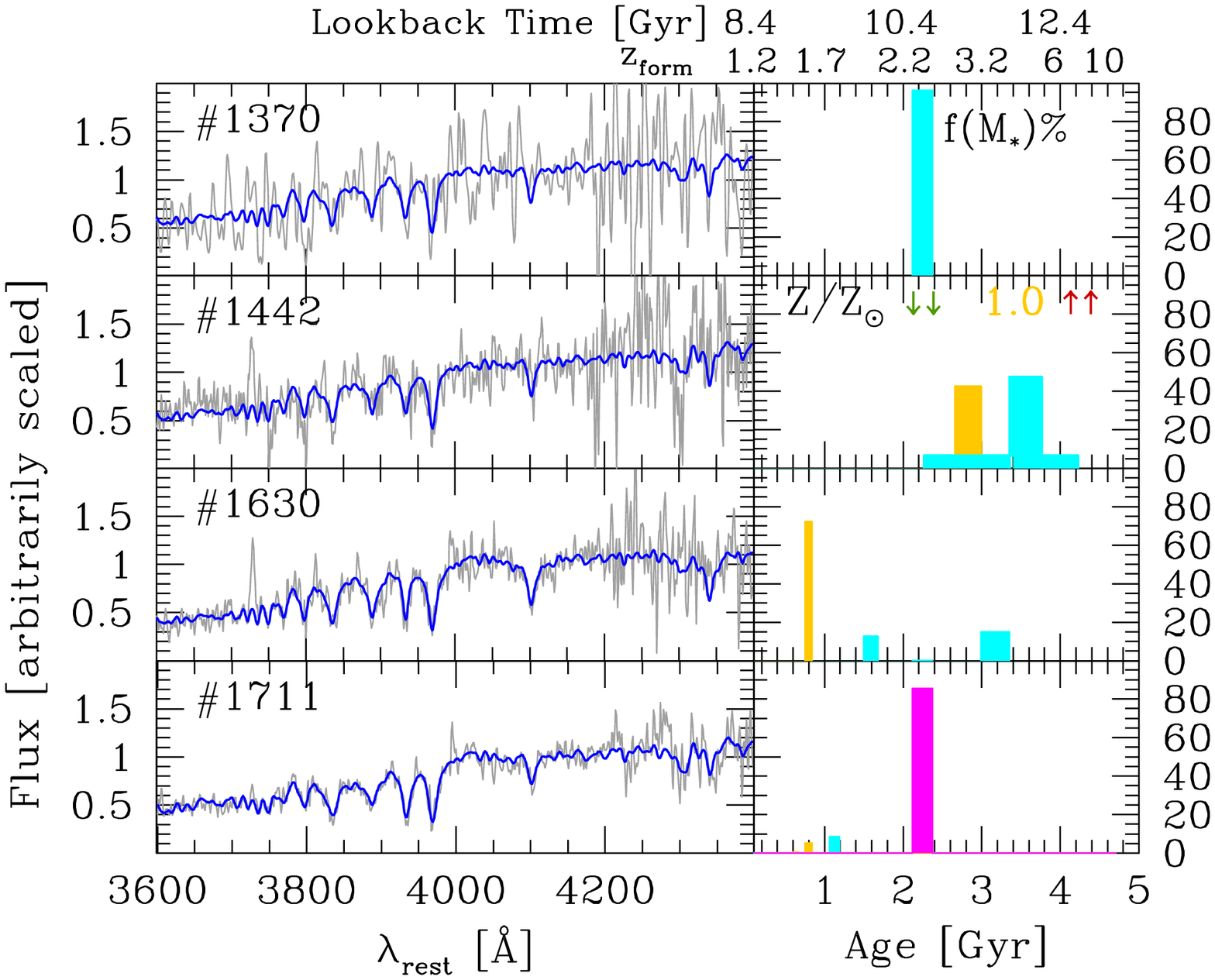}
\caption{\label{fig:star2} Results of the spectral synthesis for galaxies 
 \#1370, \#1442 and \#1630 and \#1711.  
}  
\end{figure*}
\renewcommand{\thefigure}{\arabic{figure}}

In Tab. \ref{tab:fitting} the luminosity-weighted and the mass-weighted
age and metallicity values resulting from the spectral fitting 
are reported for each galaxy, {  together with the reduced $\chi^2$ of
the best-fitting model}. 
The values refer to results obtained with the set of SSPs based on PADOVA 
isochrones.
The corresponding best-fitting
synthetic spectra are shown in Figure \ref{fig:star1}.
Results based on BaSTI isochrones are similar to those obtained with PADOVA models, 
and are not shown here for brevity reasons, with differences in age and metallicity 
smaller than 20\%, and 0.15dex, respectively.
The two extinction laws produce negligible differences.
The values of metallicity and age obtained through the spectral
fitting confirm those previously derived in \S 4:
galaxies \#651,  \#972, and \#1142 are the  oldest in our sample while galaxy \#1620 
is the youngest; galaxies \#1370  and \#1442  tend to  have lower
metallicities than the others.

The errors on age and metallicity are
the confidence intervals at 68\% confidence level of the solutions in the Age-[Z/H] space.
They have been obtained considering all the fitting 
solutions within 1$\sigma$ from the best-fitting one. 
To this end, we compared the reduced-$\chi^2$ of all the solutions 
with the reduced-$\chi^2$ of the best-fitting one making use of the F-test.
Due to the fact that errors on age and metallicity are anti-correlated,
 \citep[see, e.g.,][]{thomas05}, the confidence intervals 
should be read jointly in the sense that, older (higher) values of age (metallicity) 
correspond to lower (younger) values of metallicity (age) with respect to the best-fitting values.
{  It is worth noting that predictions of MILES-SSPs for a certain range of very young ages and 
very low metallicities, (age$<$0.1 Gyr and [Z/H]<-1.5) are flagged as 
unsafe\footnote{http://www.iac.es/proyecto/miles/pages/ssp-models/safe-ranges.php} 
by \cite{vazdekis10}.
We note, however, that there are no SSPs within the unsafe range
contributing to the fitting solutions.
}

As can be seen from Tab. \ref{tab:fitting}, luminosity-weighted and mass-weighted 
metallicities do not differ significantly (considering the errors), 
since the M/L ratio plays a secondary role in the derivation of this quantity.
The best fitting metallicity values never reach the extremes of the
range considered, with the exception of the galaxy \#1142, 
close to the maximum metallicity value allowed in the reference grid of SSPs.
Finally, we emphasize that despite to the well-known age-metallicity deneracy (whereby 
the effect of an older age is alsmost exactly compensated by that of a lower metallicity 
in the spectral synthesis), our best-fitting age and metallicity are not anti-correlated,
showing again the robustness of the results.

The resulting metallicity estimates for the 7 spheroids lie within 0.2 dex from the solar value
with a median value <[Z/H]$_M$>=0.09$\pm$0.16.
This is in agreement with the results obtained in \S \ref{sec:lines} 
based on absorption line indices (see Tab. \ref{tab:parameters}).
The above results and considerations also hold for the estimates of stellar age, 
whose median value is <Age$_M$>=2.4$\pm$0.6 Gyr.

\subsection{Metallicity variation and evolution with redshift}
\begin{figure}
\includegraphics[width=9truecm]{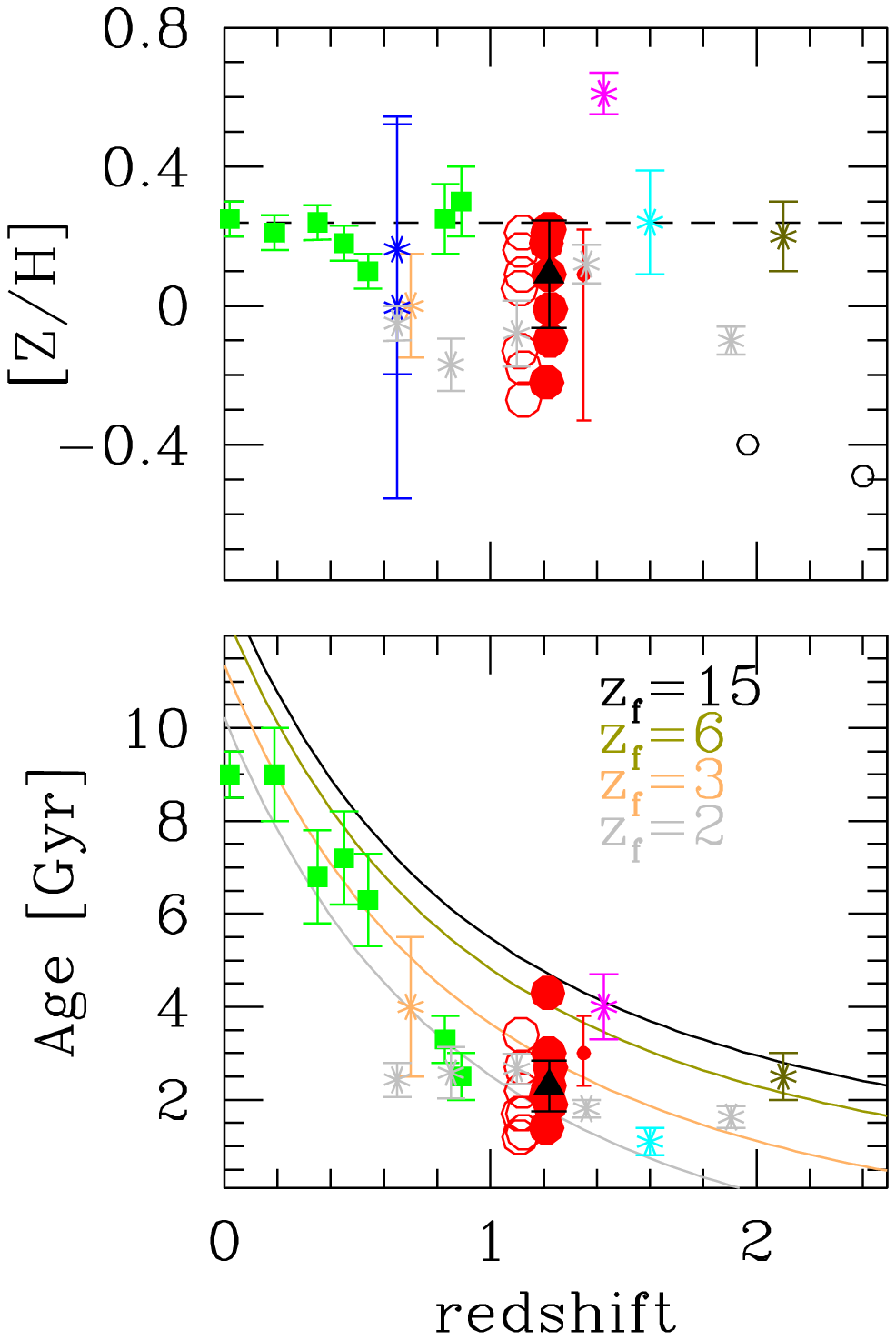}
\caption{\label{fig:metal} Stellar metallicity (upper panel) and Age (lower panel) of galaxies 
versus redshift.
The red filled circles are the mass-weighted 
values obtained for the 7 spheroidal galaxies in the cluster XLSS0223 at $z=1.22$ 
reported in Tab. \ref{tab:fitting}, the black triangle is the median value. 
The luminosity-weighted values are the open red circles offset by -0.1 in redshift 
for clarity. 
The small red dot with errorbar at $z=1.35$ shows, as example, the confidence intervals for the 
galaxy \#651.   
The green filled squares are the median values of cluster early-type galaxies 
at $0<z<0.9$ from \citet{jorgensen13} and \citet{jorgensen17}.
Skeletal symbols represent metallicity and age values of field early-type galaxies
at different redshifts from the literature: 
the blue points at $z\sim0.65$ are the median values of eBOSS galaxies with
masses log(M$_*$) in the ranges [10.7-10.9] M$_\odot$ (Z/H=-0.005) and [10.9-11.1] M$_\odot$ 
(Z/H=0.16)  \citep{comparat17},
the orange point at $z = 0.7$ is the median value of galaxies in the mass range 
10.85$<{\rm log}(\mathcal{M}_*/M_\odot)$<11 from \citet{gallazzi14}, 
the light-gray points are median values of early-type galaxies from \citet{ferreras18},
the fuchsia point at $z = 1.4$ is a massive (log$\mathcal{M}_*>11$ M$_\odot$) galaxy from
\citet{lonoce15}, 
the cyan point at $z\sim1.6$ has been derived from the stacked spectra of 24 galaxies \citep[][]{onodera15}, 
the brown point is the measurement for a massive galaxy at $z=2.1$ \citep[brown,][]{kriek16}, 
the two open circles at $z\sim2$ 
(upper panel only) are measurements of two lensed galaxies \citep[][]{morishita18}. 
The dashed line in the upper panel is the median value [Z/H]=0.24 at $0 < z < 0.9$ 
\citep[from][]{jorgensen17}.
The curves in the lower panel are predictions for passive evolution for
different values of formation redshift $z_f$, as labelled in the figure.    
}
\end{figure}

In the upper panel of Fig. \ref{fig:metal}, the metallicities of the 7 cluster
spheroids at $z=1.22$, as reported in Tab. \ref{tab:fitting}, are compared with the 
metallicities of galaxies in clusters at lower redshift and with the metallicities 
of field galaxies of comparable mass at lower and higher redshifts.
The figure shows representative median metallicity values of galaxies
in clusters in the redshift range $0.2<z<0.9$ from \cite{jorgensen17} and
\cite{jorgensen13}, the median value of field early-type galaxies  
at $z\sim0.7$ from \cite{gallazzi14} and from the eBOSS sample \citep{comparat17},
median values of early-type galaxies at different redshift bin from \cite{ferreras18},
the mean value of field galaxies at $z\sim1.6$ from \cite{onodera15} and the single measurements
of metallicity obtained by \cite{lonoce15} at $z\sim1.4$, \cite{kriek16} at $z\sim2.1$
and \cite{morishita18} at $z\sim2.2$.  

The stellar metallicity of cluster early-type galaxies remains nearly constant above 
the solar value ([Z/H]=0) up to $z\sim1.3$. 
We notice that the median value ([Z/H]$_m$=0.09) of the 7 spheroids at $z\sim1.22$ is 
slightly lower than the median value [Z/H]=0.24 estimated by \cite{jorgensen17} 
in the range $0.2<z<0.9$.
However, this difference cannot be considered as an hint for a decrease of [Z/H] with
redshift, as there is also one cluster (MS0451 at $z\sim0.54$)
in the \cite{jorgensen17} sample, whose median value of  [Z/H] ($\sim$0.1)
is fully consistent  with that we measure for galaxies in XLSSJ0223.
Thus, it might be just related to a cluster-to-cluster variation in [Z/H].

Overall, the nearly stable and constant metallicity value obtained from independent 
analysis over the redshift range 0$<$$z$$<$1.3 shows
that no significant metallicity evolution for the population
of cluster early-types has taken place over the last 9 billion years.
Thus, for the cluster spheroidal galaxies at $z\sim1.3$, 
no significant additional star formation and chemical enrichment are
required to join the present-day population of 
cluster early-type galaxies. 

Fig. \ref{fig:metal} shows also that stellar metallicity values above solar 
seem to be common in high-mass galaxies, at least up to $z\sim2$.
The figure suggests that there are no significant differences 
between the metallicity of cluster and field early-types at the redshift probed.
We notice, however, that studies of local samples do not reach 
concordant results about this issue.
Some of them find no significant differences as a function of the environment but,
rather, of the mass of galaxies, with environment
playing a role in less massive galaxies \citep[e.g.][]{bernardi09,thomas10,mcdermid15}.
On the contrary, other studies point toward significant differences, 
with galaxies in lower density environments being
less metal rich \citep[e.g.][]{cooper08,labarbera10b}. 

\subsection{Stellar age and star formation history}
The median stellar age of the seven spheroids, <Age$_M$>=2.4$\pm$0.6 Gyr, implies a 
median formation redshift $<$$z_f$$>$$\sim2.6_{-0.5}^{+0.7}$.
This value is slightly higher (not significantly, given the uncertainties) than 
the median formation  redshift ($z_f$$\sim$2) derived  
from the stellar ages and from the fundamental plane for early-type galaxies 
in clusters at $z$$<$0.9 by \cite{jorgensen17} and \cite{jorgensen13}.
It is consistent with the median formation redshift $z_f$$\sim$2.2 derived 
by \cite{woodrum17} for field early-type galaxies at $z\sim1$ with velocity 
dispersion larger 
than 170 km/s (as it is the case for six out of our seven spheroids; see Fig. \ref{fig:age_m}).

The lower panel of Fig. \ref{fig:metal} shows the stellar age of galaxies 
as a function of redshift for the same data samples considered in the upper panel.
The different curves show the prediction of stellar age for different values of 
the formation redshift in the case of passive ageing.   
The median stellar age of early-type galaxies in clusters at $z<0.9$ varies 
from cluster to cluster and, for some of them, it is consistent with formation 
redshifts even higher  than our result ($z_f$$>$3).

Contrary to metallicity, there is a significant scatter in
the ages of the seven spheroids, the largest difference being $\Delta Age\sim3$ Gyr,
independently of considering mass-weighted or light-weighted values. 
{  To estimate the significance of the scatter, we constructed N=1000 random 
realizations of the age distribution of our sample, assuming a constant mean 
age for all galaxies. 
For each galaxy, we assumed a normal distribution with lower (upper) $\sigma$ 
given by the lower (upper) relative error on its age, and the same central 
value of 2.4Gyr, the median age of our sample. 
For each realization, we estimated the standard deviation of the age 
distribution $\sigma_{age}$. 
From the distribution of $\sigma_{age}$ among different realizations, we obtained
an expected scatter of measurement $\sigma_m=0.57^{+0.33}_{-0.23}$ Gyr for 
our sample.  
Subtracting in quadrature the $\sigma_m$ to the observed standard deviation of 
age estimates for our sample ($\sim 1$~Gyr) we obtained an intrinsic age 
scatter of  $\sigma_I=0.78^{+0.24}_{-0.17}$ Gyr, 
which is significantly larger than zero at $\sim 4$ sigma level. 
}

The scatter in stellar age corresponds to a significant spread of the formation redshift
(see Tab. \ref{tab:fitting} and Fig. \ref{fig:metal}).
The seven spheroids at $z\sim1.3$ are on the ageing tracks 
that would lead them to join the present-day population of cluster early-type galaxies.
However, this does not mean that they all have to evolve passively to $z$=0.
The large scatter of their ages, indeed, leaves room to possible minor 
episodes of star formation that some of them could 
experience during the following 9 billion years.

Table \ref{tab:fitting} reports also the number of SSPs, n(SSPs), contributing to the 
synthesis of the spectrum. 
The contribution of each component, in terms of stellar mass fraction, metallicity and age, 
that is the star formation history (SFH) of each galaxy,
is shown in the right panels of Figure \ref{fig:star1}.
The histograms represent the stellar mass fraction of each component, 
the colour identifies the metallicity (green sub-solar, yellow solar, red above solar), 
the bar-code on the bottom of Fig. \ref{fig:star1} shows the ages of the SSPs
considered.
The lookback time and the formation redshift corresponding to the ages of the SSPs 
are also shown on the top x-axis of the figures. 

Given the small number of galaxies, we are not in position to study the 
dependence of the SFH of early-type galaxies at $z\sim1.3$ on galaxy mass
as done, e.g., by \cite{thomas05,thomas10}, \cite{delarosa11} or \cite{mcdermid15},
since our results would be dominated by the intrinsic scatter among the galaxies.
Hence, we consider here only the overall features of the sample.

In most cases, the bulk of the stellar mass is formed in 
a single main episode of star formation, as for galaxies \#972, \#1142, \#1370, \#1711,
and also for galaxy \#1630, the youngest galaxy most likely in quenching phase
(see \S 4.1).
Hence, these galaxies are characterized by a similar SFH.
However, their mass-weighted ages, that represent the epoch of 
formation of most of their stellar content, differ significantly,
pointing to different formation epochs.
In the remaining cases, the SFH is protracted over longer time or distributed among
different episodes of star formation, as in the case of \#651, and still not ceased,
as for \#1442.
The formation redshift of the bulk of galaxy stellar mass, is 
consistent with the mean value found by \cite{delarosa11} and \cite{mcdermid15} 
(lookback time $>$10 Gyr) for galaxies with similar dynamical mass
(log(M$_{dyn}$)$>$11 M$_\odot$), even if the SFH of some of our spheroids  
implies a larger lookback time. 

Overall, our results show that the population of massive spheroids, in this cluster at 
least, is not coeval and their stellar populations have been formed at significant 
different epochs.

\section{Age and metallicity scaling relations}

\begin{figure*}
 \includegraphics[width=15truecm]{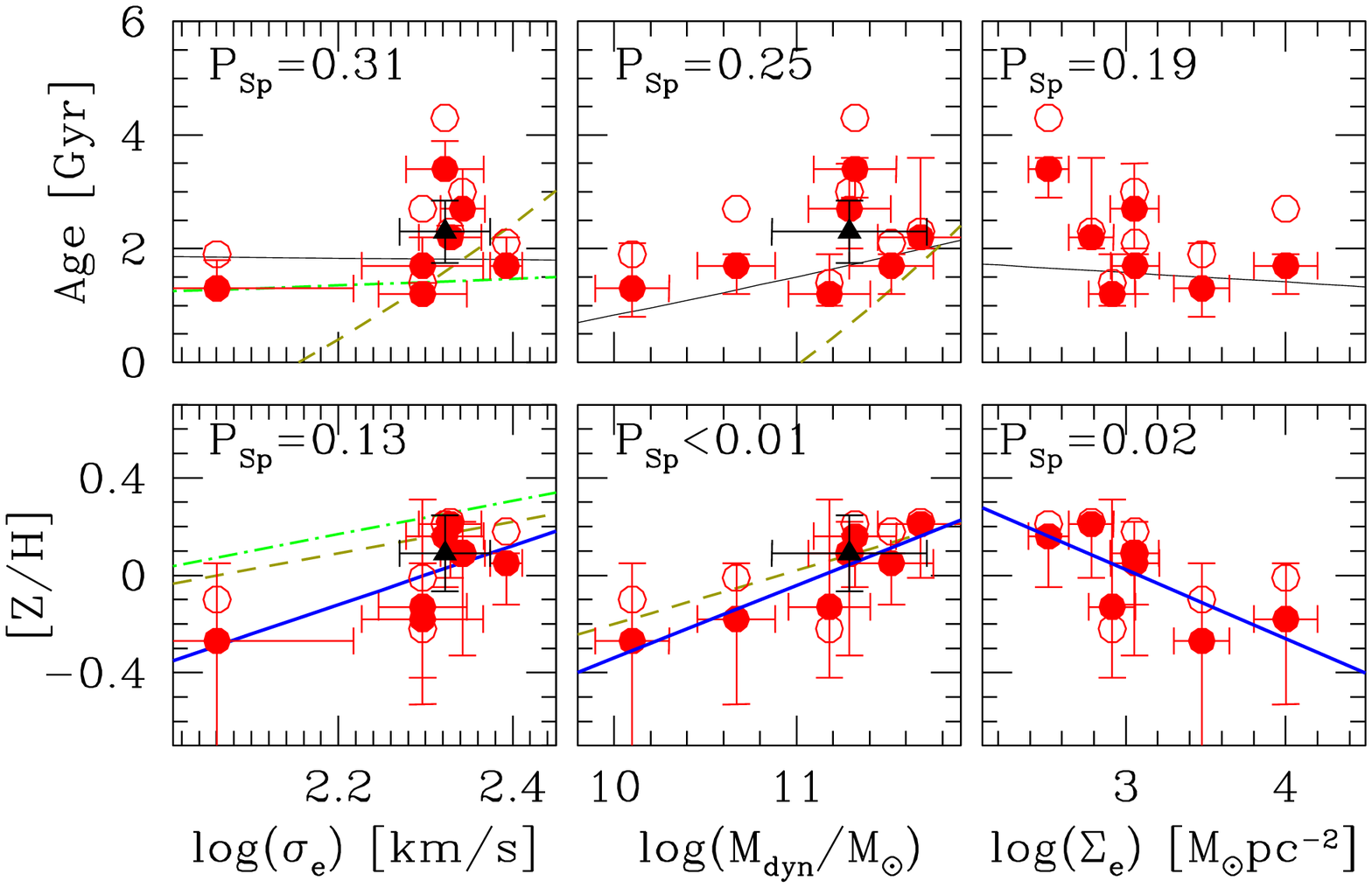}
\caption{\label{fig:age_m} Age (upper panel) and stellar metallicity (lower panel) of galaxies 
versus velocity dispersion $\sigma_e$ (left), dynamical mass M$_{dyn}$ (middle) and
stellar mass density $\Sigma_e$ (right).
Each panel reports the probability P$_{Sp}$ based on the Spearmen rank test that
the light-weighted Age and [Z/H] are uncorrelated with each quantity on the x-axis.
The red filled (open) circles are the light-weighted  (mass-weighted)
values obtained for the 7 spheroidal galaxies in the cluster XLSS0223 at $z=1.22$ 
(see Tab. \ref{tab:fitting}); the black triangle is the median value.
The dark gray dashed curves are the relations of light-weighted Age and [Z/H]
with velocity dispersion and dynamical mass found by \citet{thomas10} for local galaxies.
The black solid curves are the local relations we derived using a sampe
extrated from the SPIDER sample \citep[see \S 6;][]{labarbera10}. 
These relations are evolved back in time to $z=1.22$ in the upper panel 
under the assumption of passive evolution. 
The green dot-dashed lines are the relations found by \citet{jorgensen17}
for cluster early-types at $z$$\sim$0.9.
The thick blue lines in the lower panels are the least-squares fit
to the data.   
}
\end{figure*}

In Fig. \ref{fig:age_m}, the age and the stellar metallicity of the seven
cluster spheroidal galaxies are plotted versus their stellar velocity 
dispersion $\sigma_e$ [km $s^{-1}$], dynamical mass 
M$_{dyn}=5\sigma^2R_e/G$ [M$_\odot$] and stellar mass density within the 
effective radius $\Sigma_e=0.5M^*/\pi R_e^2$ [${\rm M_\odot\ pc^{-2}}$]. 
Velocity dispersion measurements are from Saracco et al. (2018, in preparation).
Given the small number of galaxies of our sample, it is beyond the
scope of our analysis to establish the relationships between stellar
populations properties and structural properties in early-type 
galaxies at $z$$\sim$1.3.
Our aim is to assess whether correlations can be detected in our data,
hence whether they are already present at this redshift, and how the properties
of these spheroidal galaxies compare with the local relations.

In Fig. \ref{fig:age_m}, the light-weighted parameters instead
of the mass-weighted ones,  are highlighted by red filled
circles to make the comparison with local relations, based on 
light-weighted values, easier.
The age and metallicity versus $\sigma$ relations found by \cite{jorgensen17} 
for cluster early-type galaxies in the range 0.2$<$$z$$<$0.9 are shown (green 
dot-dashed curves).
{  The local relations, age and metallicity versus $\sigma_e$ and dynamical mass,
from \cite{thomas10} (dashed curves) are also resported, for sake of completeness.
These relations, indeed, cannot be used as direct comparison for our sample,
since they have been derived from a sample that includes galaxies
at $z\sim0$ with ages younger than the lookback time ($t_{LB}\sim 8.5$~Gyr) 
corresponding to $z\sim1.2$ (especially at the lowest end of our velocity 
dispersion range), whose progenitors are likely not included in our sample.
To overcome this problem, we constructed a local sample of spheroids with properties
similar to our sample, from which we derived the local reference relations.
We extracted from the SPIDER sample \citep{labarbera10}
at $z\sim0.07$, a sample of galaxies with $\sigma_e>190$ km/s (i.e. the range
where 6 out of our 7 spheroids lie)
and ages older than 7 Gyr.\footnote{Notice that we allowed for an age
younger than $t_{LB}$, to account for the progenitor bias, 
yet leaving room to spheroids that may have experienced a secondary minor 
burst of star formation at later times.}
The derived relations, age versus $\sigma_e$, M$_{dyn}$ and $\Sigma_e$,
passively evolved back in time, are shown in Fig. \ref{fig:age_m} as black solid curves.
The stellar mass densities $\Sigma_e$ are derived using Sersi\'c effective radius and 
stellar masses based on \cite{vazdekis10} models and Chabrier IMF \cite[see][]{swindle11}.  
}

In each panel, the probability based on the Spearmen rank test that the 
two parameters shown are mutually uncorrelated is also reported. 
As expected, the relations involving the age and/or the 
velocity dispersion are those presenting the largest scatters.
We do not detect a significant correlation between age and the other
parameters of galaxies, such as $\sigma_e$, M$_{dyn}$ and $\Sigma_e$.

{  The seven spheroidal galaxies agree with what expected from the extrapolation
of the local relations, when they are passively evolved back in time.
The data at $z\sim1.2$ follow the same relations between age and mass,
velocity dispersion and mass density, than those at $z\sim0$.}
It is worth to note, that the trend of age and metallicity 
with velocity dispersion may be affected by the aperture within 
which measurements are done 
\citep[e.g.][]{labarbera10,mcdermid15}, 
because of the presence of metallicity and age gradients 
both in low-redshift \cite[e.g.][]{saglia00, wu05, labarbera09,marian18}  
and in high-redshift spheroids \cite[e.g.][]{guo11,gargiulo11,gargiulo12,chan16,ciocca17}.
However, it has been shown that measurements within different regions affect 
the offsets of the relations but not (significantly) their slope 
\citep[e.g.][]{mcdermid15,gargiulo17}.
{  The agreement of our data with the local relations and with
those found by \cite{jorgensen17} over the range $0.2<z<0.9$, confirms
this statement and the robustness of the results.}

The metallicity [Z/H] of the seven spheroids shows a positive but not statistically
significant trend with the velocity dispersion.
A least square fit to the data provides the relation
 \begin{equation}
 [Z/H]=(1.2\pm0.5) log(\sigma_e)-(2.8\pm1.2)
\end{equation}
shown in Fig. \ref{fig:age_m} (blue solid line), where $\sigma_e$ [km s$^{-1}$]
is the velocity dispersion within the effective radius R$_e$.
The slope agrees with the one found for early-type galaxies 
in the local universe \citep{thomas10} and in cluster at $z\sim0.9$
\citep{jorgensen17}, while the zero point of the relation is offset by about 
0.15 dex toward lower metallicity values.

\begin{figure}
\includegraphics[width=8truecm]{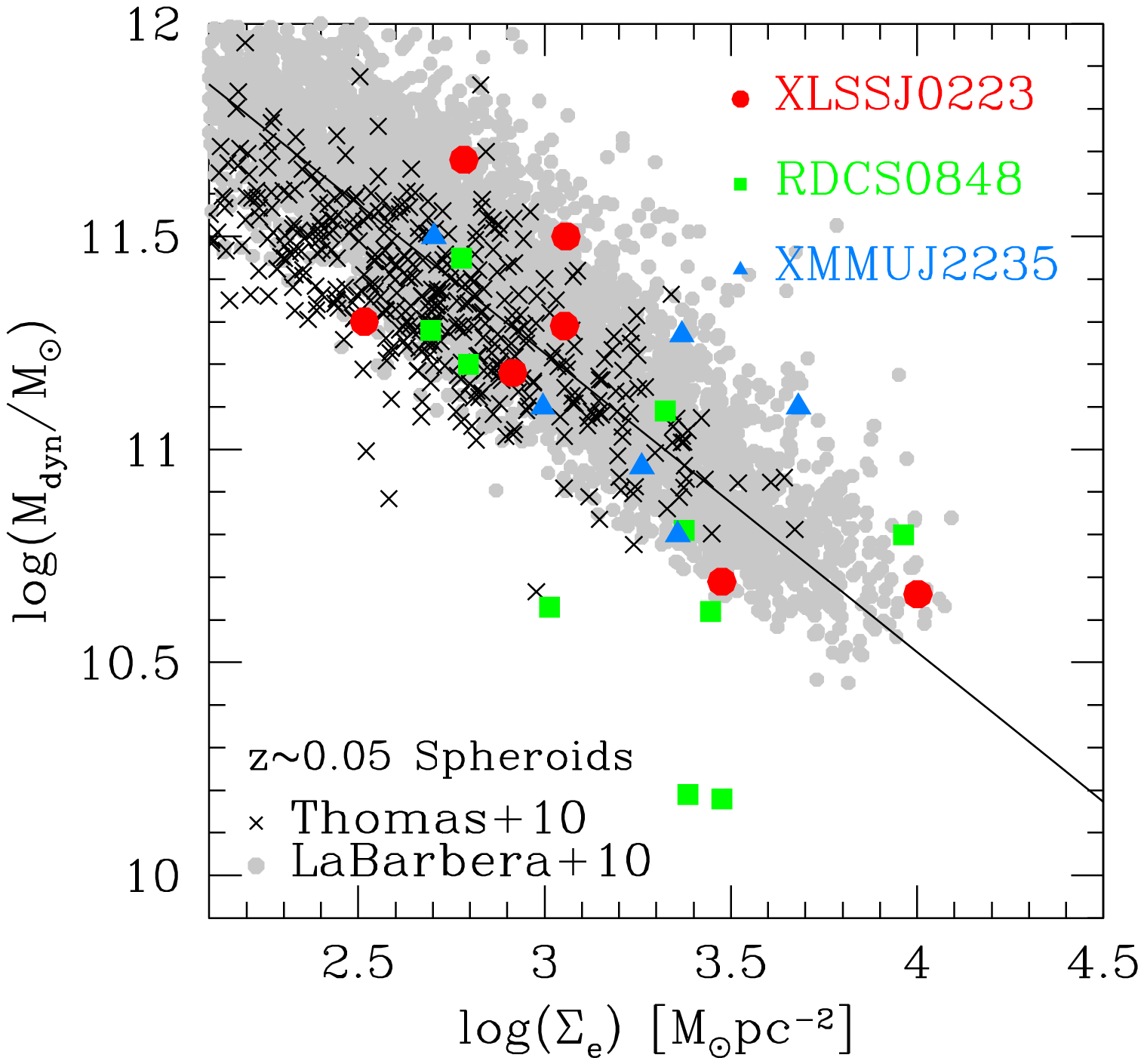}
\caption{\label{fig:mdyn_sigma} Dynamical mass versus stellar mass density
of cluster spheroidal galaxies.
The dynamical mass M$_{dyn}=5\sigma^2 R_e/G$ of early-type galaxies in 
clusters at 1.2$<$$z$$<$1.4 is plotted versus their stellar mass density
$\Sigma_e=0.5M^*/\pi R_e^2$.
The red filled circles are the 7 spheroidal galaxies in the cluster XLSSJ0223 at $z=1.22$, 
the green squares are spheroidal galaxies in the cluster RDCS0848 at $z$=1.27 from
\citet{jorgensen14}, the blue triangles are spheroidal galaxies in the cluster
XMMUJ2235 at $z$=1.39 from \citet{beifiori17}.
Black crosses are the spheroidal galaxies at $z$$\sim$0.05 
with $\sigma_e>$190 km s$^{-1}$ from the sample of \citet{thomas10}. 
Light-gray filled circles are the spheroids at similar redshift
with $\sigma_e>$190 km s$^{-1}$ extracted from the SPIDER sample \citet{labarbera10}.
The black line represents the orthogonal fit to the high-z cluster data
log(M$_{dyn}$/M$_\odot$)=(-0.7$\pm$0.2) log($\Sigma_e$)+(13.3$\pm$0.5).  
}
\end{figure}

Metallicity is correlated 
($\sim$3$\sigma$ confidence level, P$_{Sp}$=0.006) with the dynamical mass M$_{dyn}$.
The least square fit to the data provides the relation
\begin{equation}
 [Z/H]=(0.29\pm0.07) log(M_{dyn}/M_\odot)-(3.3\pm0.8)
\end{equation}
that agrees well with the local relation of \cite{thomas10}.
Metallicity shows also a negative trend with the stellar mass density 
$\Sigma_e$, with less dense galaxies being more metal rich.
These two opposite correlations are due to the anticorrelation 
between dynamical mass and stellar mass density resulting from the fundamental plane: 
galaxies with higher total mass are also less dense.
This is shown in Fig. \ref{fig:mdyn_sigma} where the dynamical mass is plotted 
versus the stellar mass density for the seven spheroidal galaxies studied here, 
belonging to the cluster XLSSJ0223 at $z$=1.22, the early-type galaxies in the 
cluster RDCSJ0848 at $z=$1.27, studied by \cite{jorgensen14} 
(samples \#4 and \#5 of their Tab. 7) and the seven spheroidal galaxies in the 
cluster XMMJ2235 at $z$=1.39 studied by \cite{beifiori17}.
Two different samples of field spheroidal galaxies in the local universe 
selected with $\sigma_e$$>$190 km s$^{-1}$, are also shown for comparison.
The first one, extracted from the sample of \cite{thomas10}, has
stellar mass densities derived using DeVaucouleur effective radii and the stellar 
masses of \cite{comparat17},
based on MS11 models \citep{maraston11} with Charbier IMF and MILES stellar library.
The second one, extracted from the SPIDER sample \citep{labarbera10}, has
stellar mass densities derived using Sersi\'c effective radius and stellar masses 
 based on \cite{vazdekis10} models and Chabrier IMF of \cite[see][]{swindle11}.  
This fundamental relation is shown here for clarity and it will be discussed in 
detail in a forthcoming paper.

\begin{figure}
 \includegraphics[width=8.5truecm]{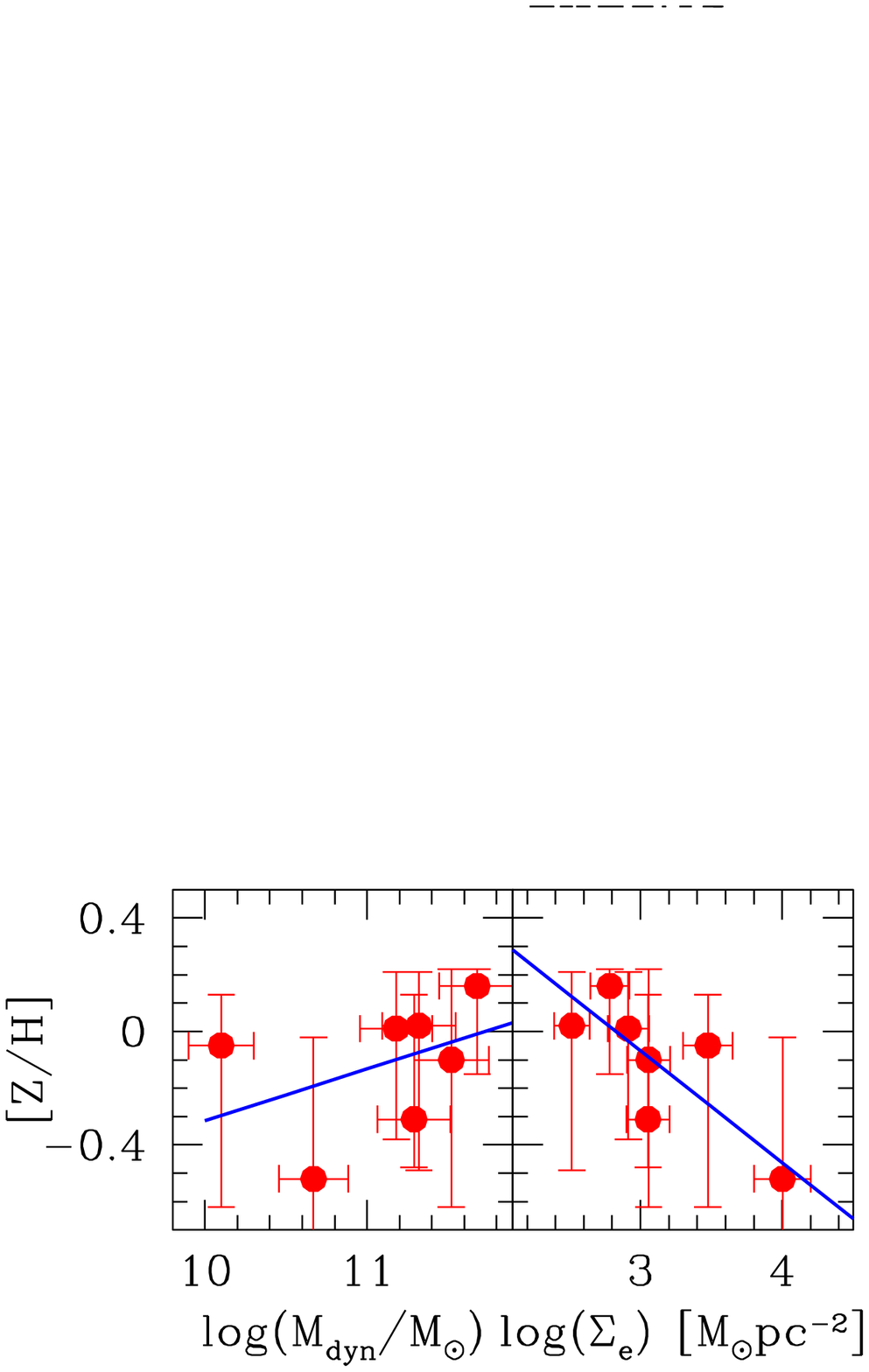}
\caption{\label{fig:labarb} Same as Fig. \ref{fig:age_m} but for
the stellar population parameters of Tab. \ref{tab:parameters}
resulting from the fitting of spectral line indices (see \S 4).
}
\end{figure}

The relationships involving metallicity, dynamical mass and 
stellar mass density, are rather robust and do not depend either on the models
used or on the fitting method.
Fig. \ref{fig:labarb} shows, as example, the relations as obtained through line strengths 
fitting reported in Tab. \ref{tab:parameters} (see \S 4).
The same results are obtained using the parameters obtained with MS11 or 
BC03 models.
Indeed, as shown in Tab. \ref{tab:bc03} and \ref{tab:m11}, within each set of results, 
galaxies keep nearly the same ranking with respect to a given parameter.

The most clear correlation we detect in our data is between metallicity 
[Z/H] and dynamical mass M$_{dyn}$.
In this case the correlation is statistically significant and becomes much 
weaker when the velocity dispersion is considered.
Overall, we find that the basic trends observed in the local universe, were already 
established at $z\sim1.3$: {  massive spheroids were characterized by higher stellar 
metallicity, lower mass density and stellar populations tendentially older than 
the lower mass spheroids.} 
It is worth to recall that, here, the mass is the total mass of the galaxy,
includes both dark and barionic matter.


\section{Summary and conclusions}
In this paper we have presented an analysis of the stellar populations
in seven spheroidal galaxies in the cluster XLSSJ0223 at $z$$\sim$1.22.
The analysis, based on our optical spectra collected at LBT, is aimed at 
constraining the epoch of formation and the evolutionary path of these seven 
spheroids through the derivation of their main stellar population parameters, 
age and metallicity, and the reconstruction of their SFHs.

We have measured absorption line strengths in the rest-frame 3400-4300 \AA\
and  used them to constrain the age and the metallicity of the stellar 
populations through the comparison with measurements at lower redshift
from the literature and predictions from stellar population models.
We find that the metallicity of these spheroids is not different from the 
metallicity measured in early-type galaxies in clusters at lower redshift.

We then derived the age and the metallicity of the stellar populations
from spectral fitting.
The resulting median metallicity of the seven galaxies is [Z/H]=0.09$\pm$0.16,
with all the values within 0.2 dex from the solar value.
These values agree with those of early-type galaxies
in clusters at 0$<$$z$<0.9 and with the few single measurements of
field early-types at higher redshift.  
The constant metallicity value  over the redshift range 0$<$$z$$<$1.3 
shows that no significant metallicity evolution for the population
of cluster early-types has taken place over the last 9 billion years.
This implies also that no significant additional star formation and chemical 
enrichment are required for these seven spheroids to join the present-day 
population of cluster early-type galaxies. 

The median mass-weighted age of the seven spheroids is <Age$_M$>=2.4$\pm$0.6 Gyr,
corresponding to a formation redshift $<$$z_f$$>$$\sim2.6_{-0.5}^{+0.7}$,
or a median Lookback Time = 11$_{-1.0}^{+0.6}$ Gyr. 
We find a significant scatter in galaxy ages and, consequently, in 
formation redshifts, showing that the population of massive spheroids, at 
least in the cluster XLSSJ0223, is not coeval.
On the other hand, galaxy ages agree with those measured for cluster early-types 
at lower redshift and passive evolution would lead our sample to join the 
present-day population.

We compared the relations between age and metallicity with velocity
dispersion and dynamical mass of the seven spheroids with those
at lower redshift and in the local universe.
We do not detect any significant correlation 
between age  and the structural parameters 
considered, velocity dispersion $\sigma_e$, dynamical mass  M$_{dyn}$ and effective 
stellar mass density $\Sigma_e$. 
This is possibly because of the large intrinsic scatter of the age
of a stellar population, which is affected by many physical/local parameters,
and the small number of galaxies.
{  The age-velocity dispersion, age-mass and age-mass density relations of the 
seven spheroids agree with those derived for local spheroids, once 
passively evolved back in time and progenitor
bias is taken into account.}

On the contrary, we find that the metallicity [Z/H]  
is correlated to dynamical mass M$_{dyn}$, according to a relation very 
similar to the one obtained for local spheroidal galaxies.
We also show that the metallicity [Z/H] is anticorrelated to the stellar mass 
density $\Sigma_e$ because of the anticorrelation between M$_{dyn}$
and  $\Sigma_e$.

The data show that no significant metallicity evolution for the population 
of cluster galaxies has taken place in the last 9 billion years.
This suggests that no additional major episodes of star formation/chemical 
enrichment or minor mergers (that would dilute the metallicity) are experienced
by cluster spheroidal galaxies at $z\sim1.3$ 
to join the local population.

We find that the basic trends observed in the local universe, were already 
established at $z\sim1.3$, with more massive spheroids having higher metallicity, 
lower stellar mass density and older ages than their lower mass counterparts.

\section*{Acknowledgements}
We thank the referee for the usefull and constructive comments 
that improved the presentation of the results.This work is based on observations carried out at the Large Binocular Telescope
(LBT) under program ID 2015\_2016\_28. 
The LBT is an international collaboration among institutions in the US, Italy 
and Germany. 
We acknowledge the support from the LBT-Italian Coordination Facility
for the execution of the observations, the data distribution and for 
support in data reduction. 
We thank D. Thomas and J. Lian for having provided us with the median 
metallicity values of eBOSS galaxies from Comparat et al. (2017).
PS would like to thank T. Durden for the useful suggestions.




\bibliographystyle{mnras}
\bibliography{paper_pop_rev} 




\appendix
\section{Results of spectral fitting with different models}

In this appendix we report the result of the spectral fitting performed
with \texttt{STARLIGHT} using a set of BC03 SSPs 
\citep{bruzual03} and a set of MS11 SSPs \citep{maraston11}.
Both the sets of models are based on Chabrier IMF.
For the BC03 models, we considered 20 ages in the range [0.06; 4.5] Gyr 
and 5 metallicities in the range [-1.7; 0.4], the results are shown in
Tab. \ref{tab:bc03}.
For MS11 models we considered 20 ages in the range [0.06; 4.5] Gyr and 5 
metallicities in the range [-2.3; 0.3], the results are shown in Tab. \ref{tab:m11}.

We emphasize that the ranking of galaxies with respect to each parameter 
(either age or metallicity), is very similar for both sets of stellar population 
models (see also \S 5), confirming the robustness of our results.

 \begin{table}
\caption{\label{tab:bc03} Age and metallicity estimates resulting from spectral fitting
with \texttt{STARLIGHT} using BC03 SSP models.}
\centerline{
\begin{tabular}{rlrrr}
\hline
\hline
  ID &  Age$_{L}$&  [Z/H]$_{L}$&  Age$_{M_*}$&  [Z/H]$_{M_*}$\\
     &   [Gyr] &     & [Gyr] &   \\
\hline
 651&   2.6  & +0.10   &2.8   & +0.19 \\  
 972&   3.2  & +0.12   &4.0   & +0.19 \\  
1142&   3.0  & +0.15   &3.6   & +0.16 \\  
1370&   2.6  & -0.40   &3.0   & -0.21 \\  
1442&   1.7  & -0.13   &2.7   & +0.01 \\  
1630&   1.0  & +0.17   &1.3   & +0.05 \\  
1711&   1.2  &  0.35   &1.9   &  0.40 \\  
\hline
\end{tabular}
}
\end{table}

 \begin{table}
\caption{\label{tab:m11} Same as Tab. \ref{tab:bc03} but for MS11 SSP models.}
\centerline{
\begin{tabular}{rlrrr}
\hline
\hline
  ID &  Age$_{L}$&  [Z/H]$_{L}$&  Age$_{M_*}$&  [Z/H]$_{M_*}$\\
     &   [Gyr] &     & [Gyr] &   \\
\hline
 651&   2.2  & +0.05   &--   & -- \\  
 972&   2.5  & +0.30   &--   & -- \\  
1142&   2.0  & +0.30   &--   & -- \\  
1370&   2.6  & -0.00   &--   & -- \\  
1442&   1.7  & -0.01   &--   & -- \\  
1630&   0.7  & +0.13   &--   & -- \\  
1711&   1.2  &  0.22   &--   & -- \\  
\hline
\end{tabular}
}
\end{table}


\bsp	
\end{document}